\documentclass[sigconf,balance=false]{acmart}

	\usepackage{popets}
	\setcopyright{popets}
	\copyrightyear{2024}
	\acmYear{2024}
	\acmVolume{YYYY}
	\acmNumber{X}
	\acmDOI{XXXXXXX.XXXXXXX}
	\acmISBN{}
	\acmConference{Proceedings on Privacy Enhancing Technologies}
\usepackage{hyperref}
\usepackage[]{algorithm2e}
\usepackage{subcaption}
\usepackage{multirow}
\usepackage{multicol}
\usepackage{siunitx}
\usepackage{color}
\usepackage{soul}
\usepackage{makecell}
\usepackage{tabulary}
\usepackage{enumitem}
\usepackage{cleveref}

\setcopyright{popets}
\copyrightyear{YYYY}

\acmYear{YYYY}
\acmVolume{YYYY}
\acmNumber{X}
\acmDOI{XXXXXXX.XXXXXXX}
\acmISBN{}
\acmConference{Proceedings on Privacy Enhancing Technologies}
\newcommand\privdnn{\textsf{PrivDNN}}

\begin{document}

\title{\privdnn: A Secure Multi-Party Computation Framework for Deep Learning using Partial DNN Encryption}


\author{Liangqin Ren}
\orcid{0009-0009-7039-0933}
\affiliation{%
  \institution{The University of Kansas}
  \city{}
  \state{}
  \country{}}
\email{liangqinren@ku.edu}

\author{Zeyan Liu}
\affiliation{%
  \institution{The University of Kansas}
  \city{}
  \state{}
  \country{}}
\email{zyliu@ku.edu}

\author{Fengjun Li}
\affiliation{%
  \institution{The University of Kansas}
  \city{}
  \state{}
  \country{}}
\email{fli@ku.edu}

\author{Kaitai Liang}
\affiliation{%
 \institution{Delft University of Technology}
  \city{}
  \state{}
  \country{}}
\email{kaitai.liang@tudelft.nl}

\author{Zhu Li}
\affiliation{%
  \institution{University of Missouri--Kansas City}
  \city{}
  \state{}
  \country{}}
\email{lizhu@umkc.edu}

\author{Bo Luo}
\affiliation{%
  \institution{The University of Kansas}
  \city{}
  \state{}
  \country{}}
\email{bluo@ku.edu}


\begin{abstract}
In the past decade, we have witnessed an exponential growth of deep learning models, platforms, and applications. While existing DL applications and Machine Learning as a service (MLaaS) frameworks assume fully trusted models, the need for privacy-preserving DNN evaluation arises. In a secure multi-party computation scenario, both the model and the data are considered proprietary, i.e., the model owner does not want to reveal the highly valuable DL model to the user, while the user does not wish to disclose their private data samples either. Conventional privacy-preserving deep learning solutions ask the users to send encrypted samples to the model owners, who must handle the heavy lifting of ciphertext-domain computation with homomorphic encryption. In this paper, we present a novel solution, namely, \privdnn, which (1) offloads the computation to the user side by sharing an encrypted deep learning model with them, (2) significantly improves the efficiency of DNN evaluation using partial DNN encryption, (3) ensures model accuracy and model privacy using a core neuron selection and encryption scheme. Experimental results show that \privdnn~reduces privacy-preserving DNN inference time and memory requirement by up to 97\% while maintaining model performance and privacy. Code is available at \url{https://github.com/LiangqinRen/PrivDNN}
\end{abstract}


\maketitle

\section{Introduction}

In the past decade, machine learning, especially deep learning \cite{DNN}, has empowered various applications, including image classification \cite{image_classification}, object detection \cite{object_detection}, and image segmentation \cite{image_segmentation}, amongst others. Deep learning relies on complex models, such as deep neural networks (DNNs), that capture specific features from input data to perform these tasks. However, developing a large/complex, and highly accurate network requires enormous training data and computing resources. While smaller organizations, such as regional hospitals and local schools, cannot afford to train their models, they expect to use the advanced models the big companies provide as a service. When it applies to sensitive information such as public security \cite{public_security}, medical data \cite{medical,ker2017deep,cancer_detection}, self-driving \cite{self_driving}, manufacturing \cite{wang2018deep}, or financial data \cite{financial}, we face a privacy dilemma in that the user data must be revealed to the entity that evaluates with the model, or the model itself must be revealed to the user to evaluate locally. In these applications with sensitive data, we should protect data and DNN model privacy while making accurate predictions, which can be treated as a \textit{secure two-party computation problem}.

Homomorphic encryption (HE) is a modern cryptographic technique based on lattices \cite{homomorphic_encryption_definition}. Fully homomorphic encryption (FHE) is the cryptosystem that supports addition and multiplication operations over encrypted data. Therefore, it is adopted in privacy-preserving outsourced storage and computation. Several works in past years introduced homomorphic encryption into deep neural networks, e.g., CryptoNets \cite{cryptonets}. In privacy-preserving DNN model evaluation, users employ FHE to encrypt the data feed to the neural network so that the model owners perform inference over encrypted data, i.e., the plaintext raw data is hidden from the model owners. On the other hand, deep learning models trained from very large amounts of proprietary and potentially sensitive data also deserve protection. \cite{first_protect_model} is the first work that considers the protection of the model, and \cite{first_protect_cnn_model} is the first work that protects convolutional neural networks. They employ the same mechanism for data protection to protect the models, i.e., use FHE to encrypt the model and feed the encrypted model with plaintext data, which will be converted into the ciphertext domain during DNN evaluation. As a result, the evaluation result is encrypted with homomorphic encryption, and only the authorized user with the key (i.e., the model owner in this case) can decrypt it and recover the plain result.

However, the homomorphic operations are extremely inefficient compared with operations in the plaintext domain. Computational overhead became the main obstacle that limits its wide deployment. For example, CryptoNets requires 250 seconds to evaluate a small self-defined FHE-friendly model on encrypted samples from the MNIST dataset with an accuracy of 99\%. Therefore, many works focus on improving the homomorphic encryption performance to make it more practical, such as \cite{cryptodl,hcnn,low_latency}.

In the applications where DNN models, instead of testing samples, are being protected, i.e., when the DNN models are shipped to the data owners to be executed locally, we do not necessarily have to encrypt the \textit{entire} model. In this paper, we present the \privdnn~framework for DNN model protection, in which we encrypt a subset of the neurons/filters to ensure that the DNN would not provide a satisfactory performance without those protected neurons. In particular, the \privdnn~scheme identifies and protects (encrypts) those ``important'' attributes that significantly contribute to the performance of the target DNN. At the inference time, since the majority of the neurons that remain in the plaintext domain do not need to be evaluated with homomorphic operations, we expect to accelerate the evaluation speed dramatically. Moreover, we do not need to use the FHE-compatible activation functions for those plaintext attributes, which will also increase the model accuracy in comparison with the approaches that employ FHE-friendly activation functions for the entire model. In this paper, we present the design of the \privdnn~framework, with the algorithms for \textit{core neuron selection}, i.e., to identify a small subset of neurons that ensures high performance for authorized DNN evaluation while significantly reducing DNN performance when it is evaluated without the contribution of the core neurons. Experiments with five popular benchmarking datasets and various DNN structures show that \privdnn~achieves the usability and security objectives while reducing private DNN evaluation time by order of magnitude compared to full-network-encryption approaches. Finally, we would like to highlight that the key idea of \privdnn, to reduce the FHE operations in DNN evaluation through partial DNN encryption, does not rely on any specific homomorphic encryption algorithms, homomorphic operations, or DNN architecture. Therefore, it could be adopted in any FHE-enabled DNN model encryption scheme to significantly improve its efficiency. In summary, the main contributions of this paper are three-fold.  

\begin{enumerate}[itemsep=0.5pt,topsep=0.5pt,leftmargin=*]
  \item We articulate the different models for privacy-preserving DNN evaluation and elaborate on the necessity for the model encryption approaches. 
  \item We present the first practical system \privdnn\ that protects deep neural network models with homomorphic encryption, which achieves sufficient protection while dramatically reducing the computation to enable efficient privacy-preserving DNN evaluation over large DNN models.
  \item We design and compare three different schemes for core neuron selection. We further demonstrate the effectiveness of the \privdnn~approach through extensive experiments. 
\end{enumerate}

The rest of the paper is organized as follows: We introduce the background and preliminaries in Section \ref{sec:preliminaries}, followed by the formal problem statement and threat model in Section \ref{sec:problem}. We present the technical details of \privdnn, the experimental results, and the security analysis in Sections \ref{sec:approach}, \ref{sec:experiments}, and \ref{sec:securityanalysis}. We summarize the literature in Section \ref{sec:relworks} and finally conclude the paper in Section \ref{sec:conclusion}.

\section{Preliminaries}\label{sec:preliminaries}

\noindent\textbf{SMC and Homomorphic Encryption.} Secure multiparty computation (SMC) enables mutually untrusted parties to jointly perform computing tasks without disclosing their private inputs \cite{MPC}. SMC was first introduced by Yao in The Millionaires' problem \cite{millionair}.  More recently, it has been introduced to new applications such as secure cloud computing and secure machine learning \cite{MPC_application}. 

Homomorphic encryption (HE) allows users to perform computations on encrypted data without first decrypting it. It has been adopted in SMC applications, including privacy-preserving machine learning (PPML) \cite{PPML}. HE schemes are roughly categorized into: \textit{partial homomorphic encryption}, \textit{somewhat or leveled homomorphic encryption}, and \textit{fully homomorphic encryption} (FHE). FHE~\cite{SMC_FHE_theory,SMC_FHE_implement} supports unlimited additions and multiplications in the ciphertext domain but requires excessive computation. There are three popular FHE schemes: \textit{Brakerski/Fan-Vercauteren (BFV)} \cite{BFV}, \textit{Brakerski-Gentry-Vaikuntanathan (BGV)} \cite{BGV} and \textit{Cheon, Kim, Kim and Song (CKKS)} \cite{CKKS}. BFV and BGV support accurate calculations over integers, while CKKS supports calculations over floats with errors. 

\noindent\textbf{Deep Learning.} A typical DNN includes an input, output, and several hidden layers, such as:

\noindent\textit{(1) The convolution layer} multiplies the input by a vector of weights and sums the results. The weights are calculated during training. The operations in the convolution layers are supported in FHE. 

\noindent\textit{(2) The pooling layer} performs downsampling by dividing the input into pooling regions and computing each region's maximum or average value. Since homomorphic encryption only supports addition and multiplication operations, the FHE-supported DNNs always use average pooling instead of maximum pooling.

\noindent\textit{(3) The activation layer} contains nonlinear functions to enable DNNs to learn complex patterns. The common activation functions include Sigmoid \cite{Sigmoid}, Tanh \cite{tanh}, ReLU \cite{ReLU}, and Swish \cite{Swish}. To implement non-linear activation in FHE, we can use polynomial substitution \cite{cryptonets}, precompute function for discrete values \cite{precompute,precompute_discrete}, use low-degree polynomials to approximate the non-linear functions \cite{cryptodl,immitate1,immitate2,immitate3}, or implement nonlinear functions through sign function \cite{sign_function}.

\noindent\textit{(4) The batch normal (BN) layer} reduces internal covariate shift by performing the normalization for each training mini-batch to allow higher learning rates. The BN layer supports FHE by using the average and variance calculated in the training process \cite{batch_normalization_fhe}.

A large DL model may have millions of parameters that require excessive computing power, memory, and storage \cite{pruning_over_parameter}.
Network pruning \cite{pruning} has been identified as an effective technique to improve efficiency. A typical pruning process includes three steps: train a large over-parameterized model, prune the model, and fine-tune the pruned model to regain the lost performance \cite{pruning_survey}. Pruning can be categorized into \textit{unstructured pruning} \cite{unstructured_pruning}, which removes specific weights of neural networks to achieve a high compression rate, and \textit{structured pruning} \cite{PFEC} pruning, which removes entire filters to accelerate with standard hardware and libraries.

\section{Problem Statement and Attack Model}\label{sec:problem}

\begin{figure*}[t]
    \centerline{\includegraphics[width=0.98\textwidth]{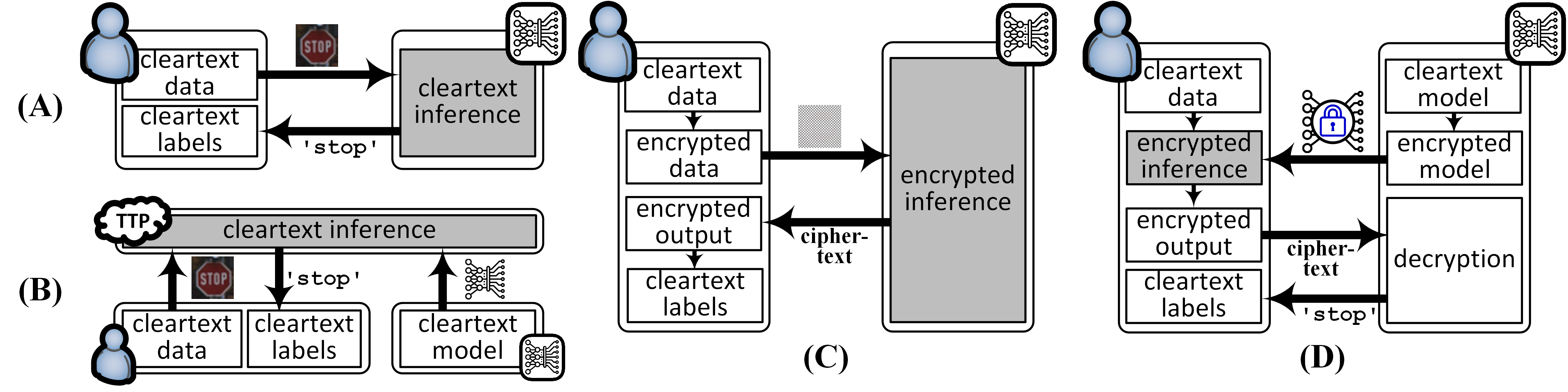}}\vspace{-3mm}
    \caption{Approaches for privacy-preserving DNN evaluation: (A) clear text approach (no privacy protection); (B) trusted third party approach; (C) data encryption approach; and (D) model encryption approach. }
    \label{fig:modes}\vspace{-2mm}
\end{figure*}

\subsection{Background and SMC Models for Privacy-preserving DNN Evaluation}

We assume a typical client-server scenario for DNN evaluation. The Server, $\mathcal{S}$, is the owner of a proprietary DNN, which is trained on private data belonging to $\mathcal{S}$. The Client, $\mathcal{C}$, is the user who has private data samples that need to be evaluated by $\mathcal{S}$'s DNN. In the conventional client-server DNN evaluation scenario, as shown in Figure~\ref{fig:modes} (A), $\mathcal{C}$ ships her raw data to $\mathcal{S}$, who performs inference in cleartext. While this approach is the most efficient, it reveals the client's data to the server, which could be undesirable or impossible in specific applications due to business, ethical, or legal considerations. For instance, consider a small regional hospital that wants to employ deep learning to predict patients' cancer categories to prepare for subsequent treatment. Due to its limited samples, it is impractical for the hospital to train its own DL model. The hospital may employ high-performance models trained by large organizations, such as Merative. However, the hospital may not be able to ship raw data samples to the service provider due to legal requirements regarding the confidentiality of the patient, such as HIPAA~\cite{HIPAA}. Meanwhile, the owner of the large DL models is unwilling to freely share the trained model with the community due to its business interests: the models are trained with large amounts of proprietary data and extensive computing resources, both of which could be highly valuable. In this scenario, a secure multi-party computing (SMC) mechanism is desired, and two types of privacy are considered: \textit{DNN model privacy} and \textit{data privacy}. 

\noindent$\bullet$~~\textbf{DNN Model Privacy.} From the model owners' perspective, they would like to maintain the ownership of the high-performance DL model. That is, the users should not be able to replicate or use the model without the owners' permission. 

\noindent$\bullet$~~\textbf{Data Privacy. } From the users' perspective, they will not reveal the raw data to the model owners or any (untrusted) third parties.

\vspace{1mm}
As shown in Figure \ref{fig:modes}, there exist four approaches for the client-server DNN evaluation problem. Here, we briefly describe each approach and discuss its advantages and disadvantages. 

\noindent\textbf{1. The Clear Text Approach. } As shown in Figure \ref{fig:modes} (A), $\mathcal{C}$ sends the plaintext data to $\mathcal{S}$, who performs inference in plaintext. \textit{Advantages}: This approach preserves DNN model privacy since the model never leaves its owner. It is very efficient since all computations are in plaintext. \textit{Disadvantages}: This approach does not preserve the users' data privacy, i.e., all data samples are revealed to $\mathcal{S}$. An alternate approach is to send the model to $\mathcal{C}$ to perform inference, which preserves the user's data but not the owner's DNN model. 

\noindent\textbf{2. The Trusted Third-Party Approach.} As shown in Figure \ref{fig:modes} (B), instead of having $\mathcal{S}$ and $\mathcal{C}$ trusting each other, they identify a third party that is trusted by both of them. The DNN model and the data are shipped in plaintext to the third party, who performs inference. \textit{Advantages}: This approach is easy to implement and fast since the inference phase is performed in plaintext. \textit{Disadvantages}: Identifying a trusted third party could be challenging or even impractical due to legal requirements. 

\noindent\textbf{3. The Data Encryption Approach.} Privacy-preserving DNN evaluation mechanisms have been proposed to perform inference on encrypted data \cite{cryptonets}. As shown in Figure \ref{fig:modes} (C), $\mathcal{C}$ encrypts her testing samples using homomorphic encryption and ships the encrypted data to $\mathcal{S}$, who performs inference in the ciphertext. The model output is returned to the user in ciphertext, who decrypts the data to retrieve the label. \textit{Advantage:} $\mathcal{S}$ and $\mathcal{C}$ do not trust each other while both model privacy and data privacy are preserved. Privacy protection is enforced by cryptographic properties rather than trust. \textit{Disadvantage:} The computation overhead is significant since the inference is performed in the ciphertext. Moreover, all the excessive ciphertext computations are performed by $\mathcal{S}$, who may not have the incentive to do so since the primary beneficiary is $\mathcal{C}$. 

\noindent\textbf{4. The Model Encryption Approach.} As shown in Figure \ref{fig:modes} (D), $\mathcal{S}$ encrypts the plaintext model with homomorphic encryption and sends the encrypted model to $\mathcal{C}$, who performs model evaluation in the ciphertext domain. The encrypted model output is returned to $\mathcal{S}$, which decrypts the data and returns the cleartext label. \textit{Advantages:} Both DNN model privacy and data privacy are perfectly preserved through cryptography (like the data encryption approach). The heavy-lifting computation is moved to $\mathcal{C}$, who should be willing to handle the overhead while enjoying the benefit of data privacy. \textit{Disadvantages:} The computation overhead is significant. Compared with the data encryption approach, an additional one-way network transmission is required for each inference. 

\subsection{The Threat Model}\label{subsec:threat_model}

We adopt the \textit{model encryption approach} (Figure \ref{fig:modes} (D))  for privacy-preserving DNN evaluation. Our scenario involves two parties: the model owner (server $\mathcal{S}$) and the data owner (client $\mathcal{C}$). We assume all participants to follow the honest-but-curious (semi-honest) model \cite{goldreich2009foundations,yao_hbc}, i.e., they precisely follow the protocol (honest), while they also actively attempt to obtain or infer knowledge about others (curious). In particular, we make the following assumptions: 

\noindent\textbf{1.} The data owner $\mathcal{C}$ knows the architecture of the model and a portion of the parameters transmitted to her in plaintext. 

\noindent\textbf{2.} The data owner cannot break the encryption to learn the encrypted parameters of the DNN model. We employ a SOTA FHE scheme, CKKS \cite{CKKS}, to encrypt DNN parameters. We assume that both the CKKS scheme and its implementation are secure. That is, only the model owner $\mathcal{S}$ is capable of decrypting any ciphertext. 

\noindent\textbf{3.} The data owner does not have $\mathcal{S}$'s proprietary data used to train the model. Otherwise, she could train her model from scratch.  

\noindent\textbf{4.} The security of the computing platforms and the communication channels are considered outside of the scope of this work.

Finally, the \privdnn~framework is designed to be adaptable for all models and FHE-encryption algorithms. That is, its functionality does not rely on any specific feature of a deep learning model, e.g., it does not assume the existence of a certain type of neurons, nor does it prohibit any neurons in the target DNN model.

\section{The \privdnn~Approach}\label{sec:approach}

\subsection{\privdnn~Overview}\label{subsec:privdnnoverview}

\begin{figure}[t]
    \centerline{\includegraphics[width=0.93\columnwidth]{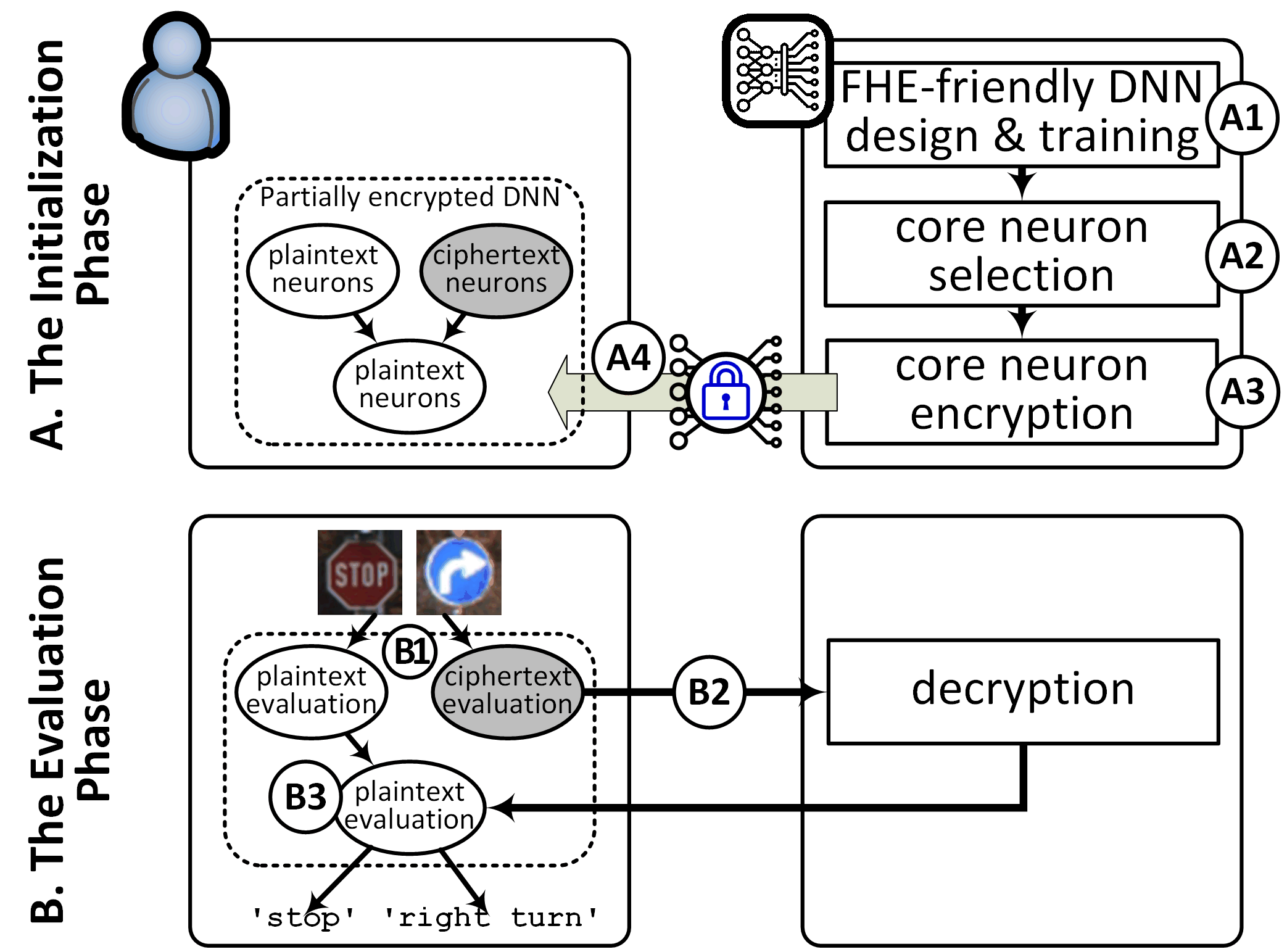}}\vspace{-3mm}
    \caption{An overview of the \privdnn~approach.}
    \label{fig:framework}\vspace{-3mm}
\end{figure}

An overview of the proposed \privdnn~approach is shown in Figure~\ref{fig:framework}. As we have described in the system model, the privacy-preserving DNN evaluation system involves two parties: the model owner ($\mathcal{S}$) and the data owner ($\mathcal{C}$). \privdnn~contains two stages: the initialization phase and the evaluation phase. 

The \textit{initialization phase} is a one-time process, which includes model training and protection: (A1) The model owner $\mathcal{S}$ designs an FHE-friendly DNN and trains it with her proprietary data. The model owner may also convert a pre-trained DNN to an FHE-friendly version with a small performance penalty. (A2) $\mathcal{S}$ identifies a set of \textit{core neurons} that are essential to the performance of the network (to be articulated in Section \ref{subsec:coreselection}). (A3) $\mathcal{S}$ employs a homomorphic encryption scheme to encrypt the core neurons. Finally, (A4) the partially encrypted DNN, which contains both plaintext and ciphertext neurons, is sent to the data owner $\mathcal{C}$.

The {\em evaluation phase} is invoked each time a batch of testing samples is evaluated by the private DNN: (B1) The data owner $\mathcal{C}$ evaluates the testing samples through the partially encrypted layers. Evaluation with the unencrypted neurons is performed in plaintext. In contrast, computation with the encrypted core neurons is performed in ciphertext using FHE. (B2) The output from the encrypted neurons will be obfuscated (multiplied by a random value), shipped to $\mathcal{S}$ for decryption, sent back to $\mathcal{C}$, and deobfuscated. (B3) With the output from the plaintext neurons and the decrypted output from the core neurons, $\mathcal{C}$ continues to evaluate the testing sample with the remaining plaintext layers of the DNN, to obtain the final output, i.e., the label for the testing sample.

In this section, we describe each component in both the initialization and the evaluation phases and articulate the technical details of the core neuron selection and partial DNN encryption mechanisms. 

\vspace{2mm}\noindent\textbf{Notations.} A pre-trained DNN model has a set of $L$ convolutional layers and $N$ neurons, among which $L^i$ is the $i$-th convolutional layer and $N_e$ is the number of encrypted neurons/filters. The model accuracy is $A_o$. The parameters in $L^i$ can be represented as a set of 4-D filters\footnote{In this paper, we use \textit{filters} and convolution layer \textit{neurons} interchangeably.} $W_{L^i}$ = \{$W_1^i$, $W_2^i$, ..., $W_{n_i}^i$\} $\in$ $\mathbb{R}^{n_{i-1} \times n_i \times k_i \times k_j}$, where the $j$-th filter is $w_j^i$ $\in$ $\mathbb{R}^{n_{i-1} \times k_i \times k_j}$. $n_i$ is the number of filters in $L_i$. $k_i{\times}k_j$ represents the kernel size (usually $k_i{=}k_j$). The output of the filters, i.e., feature maps, are denoted as $O^i$ = \{$o_1^i$, $o_2^i$, ..., $o_{n_i}^i$ \}, where the $j$-th feature map $o_j^i$ $\in$ $\mathbb{R}^{c \times h_i \times w_j}$ is generated by $w_j^i$. $c$ is the channel of input images. $h_i$ and $w_i$ are the height and width of the feature map. In \privdnn, we propose to encrypt a subset of the filters in the model to accelerate the ciphertext evaluation. Therefore, $W_{L^i}$ could be split into two groups, i.e., a subset to be encrypted $E_{L^i}$ = \{$w_{E_1^i}^i$, $w_{E_2^i}^i$, ..., $w_{E_{n_{i1}}^i}^i$\} and a subset in plaintext $P_{L^i}$ = \{$w_{P_1^i}^i$, $w_{P_2^i}^i$, ..., $w_{P_{n_{i2}}^i}^i$\}, where $E_j^i$ and $P_j^i$ together represent the $j$-th filter.

\subsection{FHE-Friendly DNN Design}\label{subsec:fhe-friendly}

The inference of \privdnn~requires computation in the ciphertext domain. Therefore, the operations involved in the encrypted neurons must be FHE-friendly. Existing works on FHE-friendly activation functions could be roughly grouped into three categories: (1) using a polynomial substitution \cite{cryptonets}, (2) precomputing function for discrete values \cite{precompute}, and (3) using low-degree polynomials to approximate the non-linear functions\cite{immitate2,immitate3,protect_data_with_low_CIFAR10_accuracy}. 

To partially mitigate the requirement of FHE-friendly activation functions, \privdnn~only requires the subset of neurons selected to be encrypted to be FHE-friendly. For example, when only the neurons from the first two convolutional layers are selected as the core neurons, we only need one linear activation function for the pooling layer because the data owner could decrypt the output from the second convolution layer.

\subsection{Partial DNN Encryption: Motivation and Objectives}\label{subsec:objectives}

In \privdnn, our core idea is to select a subset of \textit{core neurons} to be protected (encrypted) so that: (1) the accuracy of the deep neural network will decrease significantly if the core neurons are removed from the network, (2) only a small set of neurons are selected as core neurons so that the majority of the computation for DNN evaluation is still in plaintext. Meanwhile, since the output of an encrypted neuron will be ciphertext, we cannot feed it into a plaintext neuron unless it is decrypted. Otherwise, the neuron must be converted to ciphertext for FHE computation. Therefore, \privdnn~generates a partially encrypted DNN with the core neurons in the ciphertext world and the other neurons in the plaintext world. 

An example of the \privdnn~approach is shown in Figure \ref{fig:separate_remove}. Figure \ref{fig:separate_remove} (A) demonstrates an original deep neural network, whose accuracy is denoted as $A_{o}$. As shown in Figure \ref{fig:separate_remove} (B) and (C), we consider two operation modes for a partially encrypted DNN: 

\noindent (1) \textbf{Authorized DNN evaluation}. The FHE-encrypted neurons take inputs from both unencrypted neurons and other FHE-encrypted neurons. Inputs from unencrypted neurons will be encrypted to participate in cipher-domain computations. Since the output of FHE-encrypted neurons is all in the ciphertext domain, they cannot serve as input to unencrypted neurons. Therefore, as shown in  Figure \ref{fig:separate_remove} (B), the execution of the DNN is naturally \textit{separated} into the cleartext world (white) and the ciphertext world (red). Computation in the cleartext world is the same as regular DNN evaluation, except that it does not take input from the ciphertext world. FHE supports computation in the ciphertext world. It takes input from both the ciphertext world and the cleartext world, but it does not generate output to the cleartext world until the very last layer, where the ciphertext outputs are decrypted by $\mathcal{S}$ (shown in green in  Figure \ref{fig:separate_remove} (B)). We denote the accuracy of authorized DNN evaluation as $A_s$. Since the connections from FHE neurons to unencrypted neurons are eliminated, we expect $A_s$ to be slightly lower than $A_o$. 

\noindent (2) \textbf{Unauthorized DNN evaluation}. As shown in Figure \ref{fig:separate_remove} (C), an unauthorized user who obtained the partially encrypted DNN cannot decrypt the final output from the ciphertext world and join it with the plaintext world. Therefore, the FHE neurons are practically \textit{removed} from the DNN. We denote the testing accuracy of the unauthorized execution as $A_r$. Again, $A_r$ is expected to be much lower than $A_o$ due to the elimination of the neurons.  

Therefore, the primary challenge of the \privdnn~framework is to designate a subset of neurons in the target DNN as the \textit{core neurons} so that the following design objectives are satisfied:

\begin{figure}[t]
    \centerline{\includegraphics[width=0.88\columnwidth]{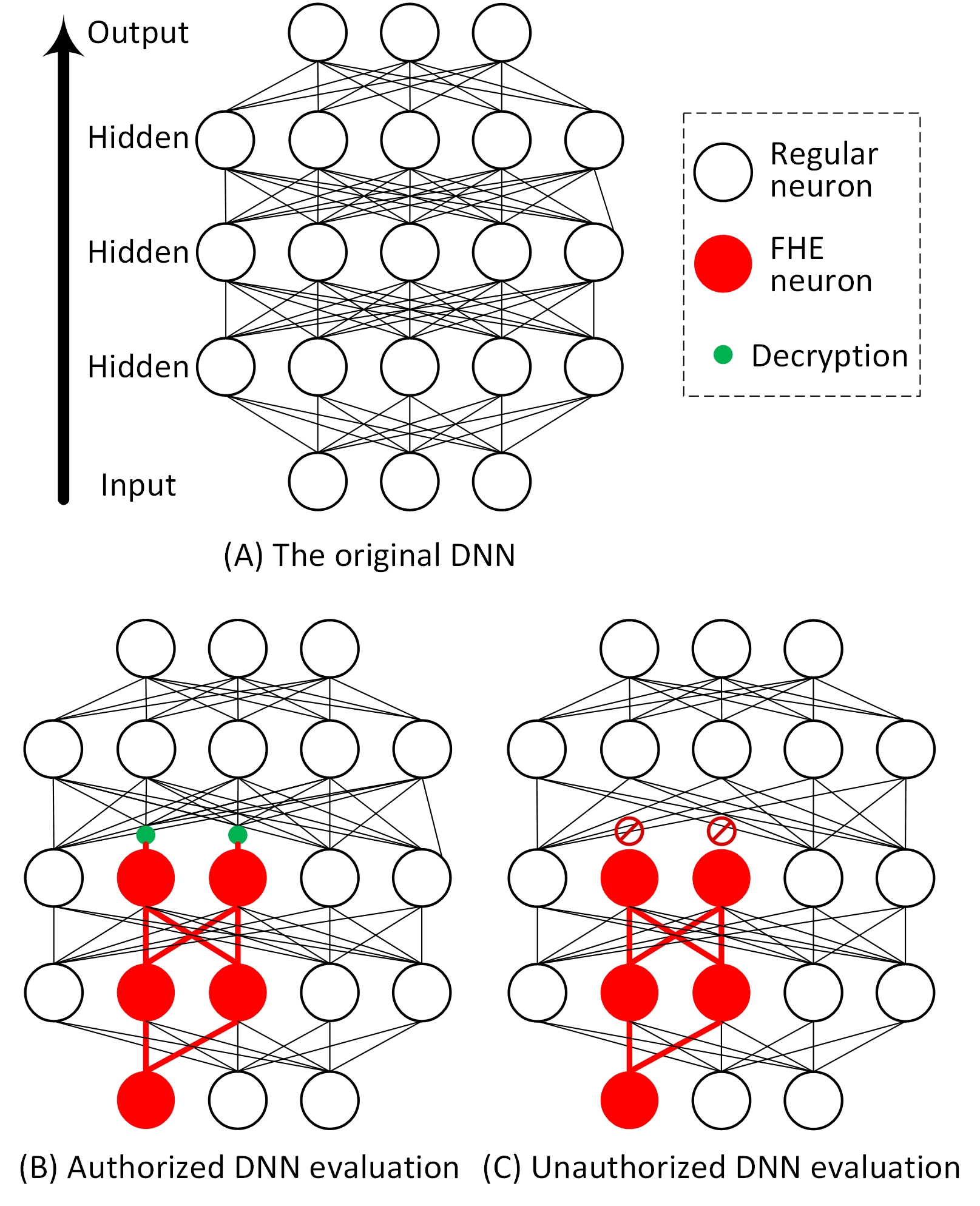}}\vspace{-3mm}
    \caption{The partial DNN encryption approach.}
    \label{fig:separate_remove}\vspace{-2mm}
\end{figure}

\noindent\textbf{G1. The Usability Goal: Maximize the Authorized Evaluation Accuracy.} The partially encrypted DNN should maintain high accuracy so that it is usable to authorized clients, i.e., the accuracy for authorized users, $A_s$, should be very close to the accuracy of the original model $A_o$. That is, $A_o-A_s$ should be close to 0.

\noindent\textbf{G2. The Security Goal: Minimize the Unauthorized Evaluation Accuracy.} The partially encrypted network is well protected against unauthorized users so that the model owner's valuable information is preserved. The model accuracy without the protected core neurons, $A_r$, should be significantly lower than the authorized evaluation accuracy $A_s$, i.e., $A_s-A_r$ should be large.

\noindent\textbf{G3. The Performance Goal: Minimize the Number of FHE-Encrypted Neurons.} The ciphertext operations are significantly slower than plaintext operations. Hence, we want to identify a relatively smaller set of core neurons so as to reduce the amount of ciphertext operations. For instance, for a CNN with $N$ convolution neurons (filters), if we identify $N_e$ filters to be encrypted, the computation of the partially encrypted model is \textit{roughly} $N_e/N$ of a full-model-encryption solution. In \privdnn, we define $N_e$ as a computation budget preset by the model owner. If the model owner has an estimation of the cost of full model encryption and chooses to accept 10\% of its cost, she would designate $N_e\simeq 0.1N$ (please see Section \ref{subsec:computationexp} for more discussions on computation). 

With a preset $N_e$, core neuron selection aims to maximize $A_s$ and minimize $A_r$ simultaneously. To evaluate whether/how a selection scheme satisfies the design objectives, we define the \textit{core set quality score} $s$ to measure the quality of the selected core neuron set:
\begin{align}
\label{eqa:total_point}
s = - sigmoid(A_o - A_s) + sigmoid(\frac{A_s - A_r}{\alpha})
\end{align}

The first component denotes a \textit{penalty} when the model accuracy decreases for authorized DNN evaluation, where {$-1<A_o-A_s<A_o<1$}. We pick the sigmoid function ($y=1/(1+e^{-x})$) to enforce a substantial penalty even for a small value of $A_o-A_s$. The second component denotes a \textit{benefit} when the model accuracy decreases for unauthorized evaluation. A higher score $s$ indicates a high-quality selection of the core neurons, i.e., a small $A_o-A_s$ for \textbf{G1} and a large $A_s-A_r$ for \textbf{G2}. In contrast, a negative $s$ indicates that the model performs badly for authorized users (large $A_o-A_s$ that outweighs the benefit of model protection). Since $A_o - A_s$ and $A_s - A_r$ have different ranges, we use a coefficient $\alpha$ as a normalization factor. In the experiments, we use an empirically selected value $\alpha=5$.

\subsection{Core Neuron Selection in \privdnn}\label{subsec:coreselection}

We present the following four algorithms for core neuron selection. In all following algorithms, we assume that the core neurons are selected from $l$ consecutive convolutional layers, where $n_{i1}$ neurons are to be selected at layer $i$, while $n_{i2}$ neurons at layer $i$ remain in the cleartext domain. $n_i$ denotes the total number of neurons at this layer so that $n_i=n_{i1}+n_{i2}$. Note that $n_{i1}$ and $n_{i2}$ are pre-selected by the model owner $\mathcal{S}$ to balance the trade-off between model security and efficiency (please see Section \ref{subsec:tradeoff} for more discussion).

For a target DNN model with $N$ convolution neurons, an exhaustive search of $N_e$ core neurons will involve testing $A_s$ and $A_r$ on $\tbinom{N}{N_e}$ different settings. While this may be feasible for small DNNs, it is impractical for any complex network. Therefore, we introduce the random selection approach as the baseline. 

\noindent\textbf{Baseline: Random Selection.}
Starting from a pre-selected layer, e.g., Layer 1, we randomly identify $n_{i1}$ neurons from layer $i$ as core neurons.  The algorithm is presented in Algorithm \ref{alg:random}. The random approach does not optimize the selection process toward the selection objectives. We introduce it as a reference to demonstrate the optimization performance of the next three approaches. 

\setlength{\textfloatsep}{1mm}
\RestyleAlgo{ruled}
\SetKwComment{Comment}{/* }{ */}
\begin{algorithm}[t]
\caption{The Algorithm for Random Selection}\label{alg:random}
\KwData{well-trained model $M$ with $L$ layers}
\KwResult{selected core neuron set $E_L$}
$i \gets GetFirstLayerIndex(M)$\;
$l \gets GetSelectLayerCount(M)$\;
\For{$i \gets i$ to $i + l$}{
    $n_{i1} \gets GetSelectFilterCount(M, i)$ \;
    $n_{i} \gets GetFilterCount(M, i)$ \;
    $E_{L^i} \gets RandomSelectFilters(n_{i}, n_{i1})$ \;
}
\end{algorithm}

\noindent\textbf{The Greedy Approach.} We aim to select core neurons that maximize $s$ as defined in Equation (\ref{eqa:total_point}). We design a greedy algorithm (Algorithm \ref{alg:greedy}) to select the neurons layer-by-layer. In layer $i$, we first temporarily add each neuron $j$ to the core neuron set (denoted as $E_{Li}$) and then evaluate the DNN performance $A_r$ and $A_s$ to calculate the core set quality score $s_j$. We select the neuron that produces the best $s_j$ and permanently add it to core neurons. We repeat the process for this layer until $n_{i1}$ neurons are added to the core set. We then move to the next layer until we finish all $l$ layers. 

The greedy selection is a dynamic evaluation approach in which the accuracy of the target DNN (both $A_s$ and $A_r$) is evaluated for different core neuron selections. Therefore, the greedy approach is computationally more expensive than static analysis, in which the core neurons are selected by analyzing the architecture and weights of the target neural network. Moreover, the greedy selection may reach a local but not the global optimum. In the experiments, we show that the greedy approach slightly compromises selection efficiency to achieve a high quality of the core neuron set. 

\begin{algorithm}[t]

\setlength{\textfloatsep}{2mm}
\caption{The Algorithm for Greedy Selection}\label{alg:greedy}
\KwData{well-trained model $M$ with $L$ layers}
\KwResult{selected core neuron set $E_L$}
$i \gets GetFirstLayerIndex(M)$\;
$l \gets GetSelectLayerCount(M)$\;
\For{$i \gets i$ to $i + l$}{
    $n_{i1} \gets GetSelectNeuronCount(M, i)$ \;
    $P_{L^i} \gets GetAllNeurons(M, i)$ \;
    $E_{L^i} \gets \{\}$ \;
    \While{$count(E_{L^i}) < n_{i1}$} {
        \ForEach{$f$ in $P_{L^i}$}{
            $s[f] \gets CalculateQuality(M, E_{L}, f)$ \;
        }
        $m \gets argmax(s[f])$ \Comment*[r]{``best'' (one) neuron}
        {\If{$s[m] \leq s[E_{L^i}[-1]]$}{ 
            break \Comment*[r]{optional}
            }
        }
        $E_{L^i} \gets E_{L^i} \cup \{m\} $ \;
        $P_{L^i} \gets P_{L^i} - \{m\} $ \;
    }
}
\end{algorithm}

\noindent\textbf{The Pruning-based Approach.} Research on DNN pruning aims to remove redundant filters in a DNN to produce sparse models for performance acceleration and model compression. The neurons/filters left after the pruning are supposed to be the ones that contribute the most to the model's performance.  Therefore, our application may adopt the pruning algorithms in the literature for core neuron selection. As shown in Algorithm~\ref{alg:pruning}, we assume that an arbitrary pruning function $Prune(M, i, n_{i2})$ is employed to identify $n_{i2}$ neurons in the $i$th layer of DNN model $M$ to be pruned. The remaining $n_{i1}$ unselected neurons are added to the core neuron set. Note that we do not perform the actual pruning operation, i.e., the ``pruned'' neurons are left intact in the model. 

\privdnn~requires the following for the pruning algorithm: (1) The algorithm should be structured pruning. Limited by FHE properties, our protection is based on the neurons (filters) rather than weights. Therefore, we can only employ structured pruning, not weight pruning. (2) The algorithm must prune a pre-trained model instead of establishing a new, smaller model from scratch \cite{rethinking_pruning, regularization}. 
(3) The algorithm should be one-shot rather than progressive \cite{movement_pruning, guide_pruning} because we cannot modify the model. 
(4) The algorithm should generate a static rather than a dynamic selection that changes based on the input. In \privdnn, the model owner $\mathcal{S}$ will share a partially encrypted model with the data owner $\mathcal{C}$. Since the model owner does not know any information about the dataset at the downstream user $\mathcal{C}$, we must use algorithms that can generate static selections rather than the dynamic pruning algorithms that change the selection based on the testing data \cite{dynamic_pruning,runtime_pruning}.

Considering the above four limitations, we select four classic pruning algorithms: PFEC \cite{PFEC}, FPGM \cite{FPGM}, HRank \cite{hrank}, and GFS \cite{GFS}. The first two are weight-dependent algorithms, and the other two are not. 
PFEC\cite{PFEC} proposes to prune filters based on their $l$-norm values. When a filter with smaller $l$-norm values multiplies the input, its output is also more likely to be smaller, i.e., its output is more likely to fail to pass sectional activation functions like ReLU. It does not contribute to the next layer. Therefore, there is unlikely any harm in pruning this filter. FPGM \cite{FPGM} argues that the norm-based criterion may not be accurate when the norm deviation is too small or the minimum norm of filters is not small. Therefore, it prunes filters with redundant information, i.e., the filters can be replaced with other filters. HRank \cite{hrank} suggests that filters with a higher rank should have more information. At the same time, the rank of a specific filter is relatively stable with different inputs. Therefore, HRank calculates filters' ranks with a small data set, and prunes filter with smaller ranks. GFS \cite{GFS} is a forward greedy algorithm. Most pruning algorithms will start with the complete model and remove filters based on their strategies. GFS removes all filters at the beginning and adds filters back individually. During this process, GFS calculates the loss that adding back a filter will decrease and selects the filter that decreases the most loss.

\setlength{\textfloatsep}{1.7mm}
\begin{algorithm}[t]
\caption{The Algorithm for Pruning-based Selection}\label{alg:pruning}
\KwData{well-trained model $M$ with $L$ layers}
\KwResult{selected core neuron set $E_L$}
$i \gets GetFirstLayerIndex(M)$\;
$l \gets GetSelectLayerCount(M)$\;
\For{$i \gets i$ to $i + l$}{
    $n_{i1} \gets GetSelectFilterCount(M, i)$ \;
    $n_{i} \gets GetFilterCount(M, i)$ \;
    $P_{L^i} \gets GetAllNeurons(M, i)$ \;
    $E_{L^i} \gets P_{L^i} - Prune(M, i, n_{i} - n_{i1}) $ \;
}
\end{algorithm}

\noindent\textbf{The Pruning+Greedy Approach}. 
Both pruning and greedy strategies have advantages and disadvantages. The greedy algorithm provides solid selection performance (will elaborate in Section \ref{sec:experiments}), but it is less efficient since it requires on-the-fly evaluation of DNN models in the selection process. The static-analysis-based pruning approaches are very fast, however, they select a subset of neurons that do not provide optimal accuracy for authorized users. Therefore, we propose integrating these two strategies as shown in Algorithm~\ref{alg:pruning_greedy}. We first employ the pruning algorithm to exclude the clearly redundant neurons, i.e., to reduce the size of the selection pool for the subsequent greedy algorithm. In practice, we create a greedy selection pool whose size is $p$ times the target core set size (we empirically set $p=2$ in our experiments). Then, we use the greedy strategy to select filters from the pool.

\begin{algorithm}[t]
\caption{Pruning+Greedy Selection}\label{alg:pruning_greedy}
\KwData{well-trained model $M$ with $L$ layers}
\KwResult{selected core neuron set $E_L$}
$i \gets GetFirstLayerIndex(M)$\;
$l \gets GetSelectLayerCount(M)$\;
\For{$i \gets i$ to $i + l$}{
    $n_{i1} \gets GetSelectFilterCount(M, i)$ \;
    $n_{i} \gets GetFilterCount(M, i)$ \;
    $P_{L^i} \gets GetAllNeurons(M, i)$ \;

    \While{$count(E_{L^i}) < n_{i1}$} {
        \ForEach{$f$ in $P_{L^i} - Prune(M, i, (n_{i} - n_{i1}) \times p)$}{
            $s[f] \gets CalculateQuality(M, E_{L}, f)$ \;   
        }
        $m \gets argmax(s[f])$ \Comment*[r]{``best'' (one) neuron} 
        {\If{$s[m] \leq s[E_{L^i}[-1]]$}{ 
            break \Comment*[r]{optional}
            }
        }
        $E_{L^i} \gets E_{L^i} \cup \{m\} $ \;
    }
}
\end{algorithm}

\section{Experiments}\label{sec:experiments}

\subsection{Settings}

We implement the \privdnn~framework with all three core neuron selection approaches (presented in Section \ref{subsec:coreselection}) using Python 3.10.13, PyTorch 2.1.0, and CUDA 12.1. All the experiments presented in this section are performed on a desktop computer with Ubuntu 22.04 LTS running on AMD Ryzen 7 3700X eight-core CPU, NVIDIA 3090 GPU, and 64 GB memory. To support DNN evaluation in the ciphertext domain, we adopt the CKKS \cite{CKKS} scheme as implemented in Microsoft SEAL 4.1.1 \cite{sealcrypto} library. As shown in Table~\ref{tab:datasets}, we adopt five popular benchmarking datasets (See Appendix~\ref{apdx:datasets} for details). 

\setlength{\textfloatsep}{20.0pt plus 2.0pt minus 4.0pt}
\begin{table}[t]
  \centering
  \caption{Datasets and models used in the experiments.}
  \label{tab:datasets}\vspace{-3mm}
  \setlength\tabcolsep{3.5pt}
  \begin{tabular}{c c c c c c}
    \hline
     & M & E & G & C & T \\
    \hline
    Train size & 60000 & 124800 & 39209 & 50000 & 100000\\
    Test size & 5000 & 10400 & 6315 & 5000 & 10000\\
    DNN Model & LeNet-5 & LeNet-5 & AlexNet & VGG16 & ResNet18 \\
    \hline
    Accuracy($A_o$) & 99.36\% & 93.08\% & 93.51\% & 90.22\% & 72.00\%\\
    \hline
  \end{tabular}
  \begin{minipage}{8.50cm}
  \vspace{0.1cm}
  \small  {\textbf{1.} M: MNIST \cite{MNIST_dataset}, E: EMNIST \cite{EMNIST}, G: GTSRB \cite{GTSRB}, C: CIFAR10 \cite{CIFAR10_dataset}, T: Tiny-ImageNet \cite{tinyimagenet} \textbf{2.} We adopt top-5 accuracy for Tiny-ImageNet.}
  \end{minipage}\vspace{-3mm}
\end{table}

As discussed in Section \ref{subsec:fhe-friendly}, \privdnn, as well as all FHE-based privacy-preserving DNN evaluation schemes in the literature, requires FHE-friendly DNN models. We use the FHE-friendly square function as the ciphertext-domain activation function. In practice, \privdnn~may train the FHE-friendly DNN from scratch or protect a pre-trained DNN. We convert the activations to FHE-friendly for pre-trained models and tune the modified model. In the experiments, we train all the models from scratch. As shown in Table \ref{tab:datasets}, the FHE-friendly models for MNIST, EMNIST, GTSRB, CIFAR10, and Tiny-ImageNet get the accuracy ($A_o$) of 99.36\%, 93.08\%, 93.51\%, 90.22\%, and 72.00\%. We focus on protecting the model rather than increasing the model's training performance or accuracy. Hence, we use the same parameters for the scheduler, batch size (128), and epochs (128) in all models.

\begin{figure}[t]
    \centerline{\includegraphics[trim={0 0 0 1cm},clip, width=1.00\columnwidth]{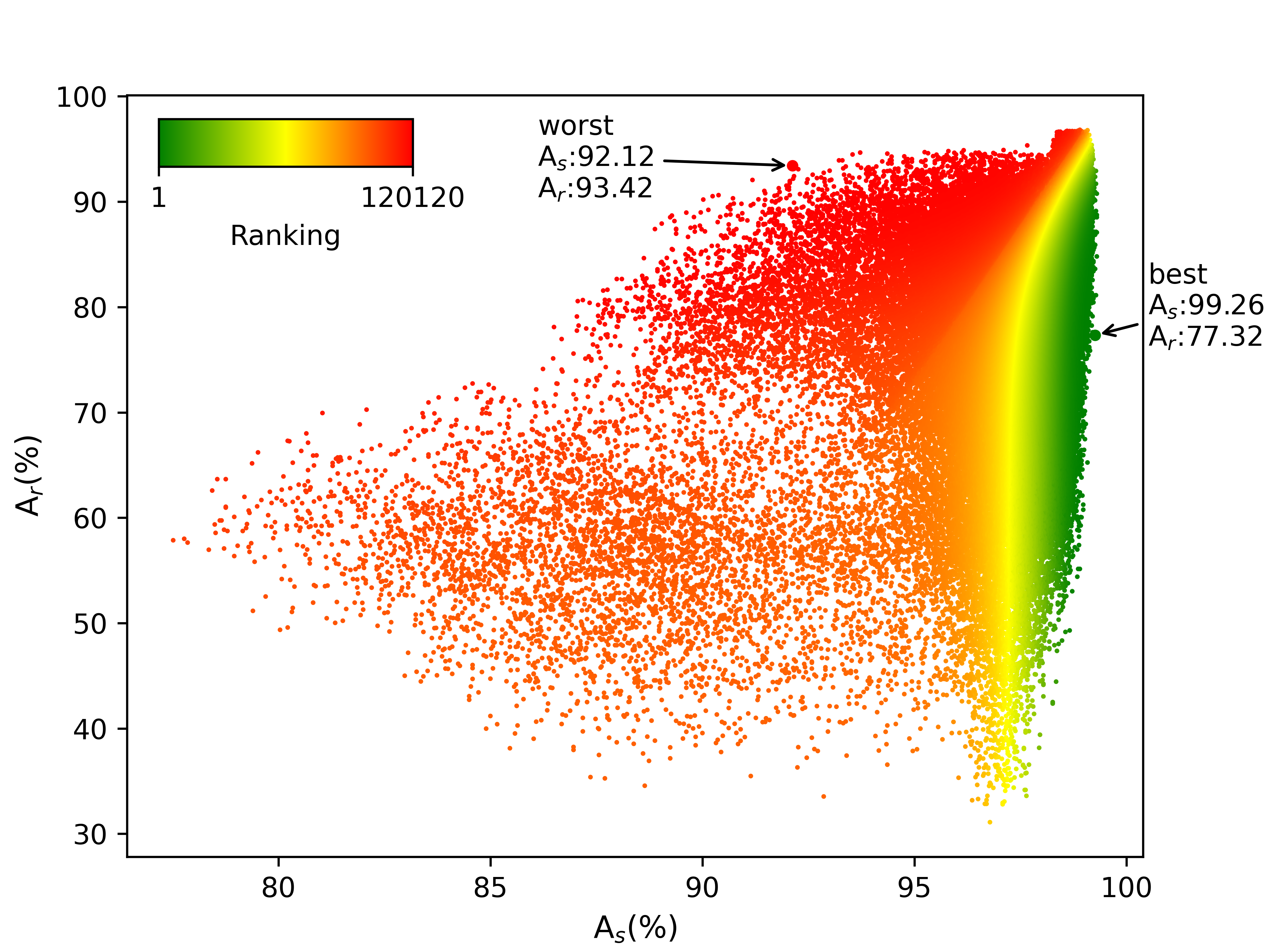}}
    \vspace{-4mm}
     \caption{The distribution of $\{A_s, A_r\}$ for all possible core neuron set selections for MNIST/LeNet-5 ($n_{1,1}=2, n_{2,1}=6$).}
    \label{fig:mnistditribution}\vspace{-3mm}
\end{figure}

\subsection{Core Neuron Selection}

\noindent\textbf{Ground Truth.} As discussed in Section \ref{subsec:coreselection}, we should perform an exhaustive search to learn the quality of all possible core neuron sets. For instance, to encrypt $n_{i1}$ neurons out of $n_i$ neurons from layer $i$, there are ${n_i \choose n_{i1}}$ different selections to be evaluated. Assume that 30\% of the neurons from the models' first two convolution layers are selected to be encrypted. Examples of possible combinations for exhaustive searches are calculated as follows:

\begin{equation}
\begin{alignedat}{3}
\label{combination}
C_{\text{LeNet-5(MNIST)}} &= \binom{6}{2} &&\times \binom{16}{5} &&= 6.522 \times 10^{4} \\
C_\text{VGG16}   &= \binom{64}{20} &&\times \binom{64}{20} &&= 3.849 \times 10^{32}
\end{alignedat}
\end{equation}

We run exhaustive searches for all possible core neuron selections for the following scenarios: (1) when \textit{up to} 2 neurons in the first layer and up to 6 neurons in the second layer of the LeNet-5 model for MNIST are selected to be encrypted; (2) when up to 2 neurons in the first layer and up to 4 neurons in the second layer of the modified LeNet-5 model for EMNIST are selected. For each $\{n_{11}, n_{21}\}$ pair, we present the number of possible selections in Table~\ref{tab:all_combinations_of_lenet5}. For each core neuron selection, we evaluate the classification accuracy of authorized and unauthorized DNN evaluation ($A_s$ and $A_r$). The best-case and the worst-case performances (in $A_s, A_r$, and $s$) are also reported in Table \ref{tab:all_combinations_of_lenet5}. For example, when 2 and 6 neurons from the first two convolution layers are selected for MNIST, there exist 120,120 possible selections. The best selection achieves a 99.26\% accuracy for \textit{authorized} DNN evaluation, i.e., a 0.1\% performance drop. Meanwhile, it achieves a 77.32\% accuracy for \textit{unauthorized} DNN evaluation, i.e., a 22.04\% performance drop.

For each $\{n_{11}, n_{21}\}$ pair in the ground truth dataset, we rank all the core neuron set selections by their quality score $s$ and use it as a reference for future experiments. We have plotted the distribution of all 120,120 $\{A_s, A_r\}$ pairs for ($n_{1,1}=2, n_{2,1}=6$) for the MNIST dataset and LeNet-5 model, as shown in Figure \ref{fig:mnistditribution}. Since we assign a higher weight to $A_s$ in the calculation of $s$, the top-ranked selections are mostly located on the right side of the plot, i.e., with higher $A_s$. A good selection algorithm is expected to find a highly ranked selection with significantly less computation than an exhaustive search. Note that we do not have the ground truth data for complex models, i.e., AlexNet, VGG16, and ResNet18, due to the excessive computation required for exhaustive searches. 

\begin{table}[t]
  \centering
  \caption{Ground truth of LeNet-5.}
  \label{tab:all_combinations_of_lenet5}\vspace{-3mm}
  \setlength\tabcolsep{4.5pt}
  \begin{tabular}{c|c|c|c|c|c|c}
    \hline
    \multirow{2}*{} & \multirow{2}*{$n_{1,1}$, $n_{2,1}$} & \multirow{2}*{Count} & \multicolumn{2}{c}{Best} \vline & \multicolumn{2}{c}{Worst} \\
    \cline{4-7}
    ~ & ~ & ~ & $s$ & $A_s$ \ \ \ $A_r$ & $s$ & $A_s$ \ \ \ $A_r$ \\
    \hline
    \multirow{10}*{\rotatebox{90}{MNIST \cite{MNIST_dataset}}} & 1, 1 & 96 & 0.16 & 99.30 95.66 & -0.63 & 93.36 96.14 \\
    ~ & 1, 2 & 720 & 0.37 & 99.30 89.30 & -0.65 & 92.12 95.16 \\
    ~ & 1, 3 & 3360 & 0.41 & 99.28 86.04 & -0.65 & 91.32 94.40 \\
    ~ & 1, 4 & 10920 & 0.44 & 99.30 83.98 & -0.64 & 92.16 95.02 \\
    ~ & 1, 5 & 26208 & 0.47 & 99.32 80.86 & -0.60 & 92.12 94.18 \\
    ~ & 1, 6 & 48048 & 0.48 & 99.34 76.64 & -0.56 & 91.74 92.86 \\
    \cline{2-7}
    ~ & 2, 2 & 1800 & 0.24 & 98.66 87.32 & -0.69 & 89.74 93.80 \\
    ~ & 2, 3 & 8400 & 0.31 & 98.86 85.36 & -0.70 & 89.50 93.76 \\
    ~ & 2, 4 & 27300 & 0.39 & 99.06 82.24 & -0.66 & 89.84 93.08 \\
    ~ & 2, 5 & 65520 & 0.42 & 99.18 82.32 & -0.60 & 90.44 92.52 \\
    ~ & 2, 6 & 120120 & 0.46 & 99.26 77.32 & -0.56 & 92.12 93.42 \\
    \hline
    \multirow{7}*{\rotatebox{90}{EMNIST \cite{EMNIST}}} & 1, 1 & 200 & 0.17 & 93.02 89.12 & -0.52 & 83.26 83.75 \\
    ~ & 1, 2 & 1900 & 0.40 & 93.02 81.37 & -0.48 & 83.85 83.38 \\
    ~ & 1, 3 & 11400 & 0.47 & 93.03 71.62 & -0.43 & 84.39 83.01 \\
    ~ & 1, 4 & 48450 & 0.49 & 93.08 68.24 & -0.36 & 84.55 81.78 \\
    \cline{2-7}
    ~ & 2, 2 & 8550 & 0.36 & 92.86 80.91 & -0.70 & 60.01 64.13 \\
    ~ & 2, 3 & 51300 & 0.43 & 92.85 70.85 & -0.65 & 59.60 62.76 \\
    ~ & 2, 4 & 218025 & 0.46 & 92.94 69.22 & -0.55 & 68.87 69.92 \\
    \hline
  \end{tabular}
  
  \begin{minipage}{8.4cm}
  \vspace{0.1cm}
  \small  Notes: \textbf{1.} $A_o$ of MNIST and EMNIST are 99.36\% and 93.08\%, respectively. \textbf{2.} Count: the number of possible combinations with given $\{n_{11}, n_{21}\}$ pair.
  \end{minipage}\vspace{-2mm}
\end{table}

\vspace{1mm}\noindent\textbf{Random and Greedy Selection.} For the datasets with ground truth selection performance, we run the random selection (Algorithm \ref{alg:random}) and the greedy selection (Algorithm \ref{alg:greedy}) algorithms and report their performance in Table \ref{tab:random_selec_of_lenet5}. For the greedy approach, we report the rank of the selected core neuron set as referenced to the ground truth results. We also report the time (in seconds) to complete the selection. As shown in the table, the greedy approach identifies the best core neuron sets in 13 out of 18 $\{n_{11}, n_{21}\}$ pairs, while the selections are very close to the best in the other five cases, e.g., ranked 3rd out of 10,920 selections (top 0.027\%).  

For each $\{n_{11}, n_{21}\}$ pair, we run the random selection algorithm for the same length of time that was used by the greedy approach, e.g., 22 seconds for $\{1, 1\}$ in MNIST. We pick the highest rank among all the random selections. We repeat this experiment 1000 times and report the average highest rank in Table \ref{tab:random_selec_of_lenet5}. Moreover, we also keep the random selection experiments running and evaluate how much time it takes for random selection to identify a core set that ranks as high as the greedy approach. The result is reported as $T_g$ in the table. As shown in the table, the greedy algorithm significantly outperforms the random selection approach. 

\begin{table}[t]
  \centering
  \caption{Random and greedy selections for LeNet-5.}
  \label{tab:random_selec_of_lenet5}\vspace{-3mm}
  \begin{tabular}{c|c|c|c|c|c|c}
    \hline
    \multirow{2}*{} & \multirow{2}*{$n_{1,1}$, $n_{2,1}$} & \multirow{2}*{Count} & \multicolumn{2}{c}{Greedy} \vline & \multicolumn{2}{c}{Random} \\
    \cline{4-7}
    ~ & ~ & ~ & \# & T & Avg(BestRank) & $T_g$ \\
    \hline
    \multirow{10}*{\rotatebox{90}{MNIST \cite{MNIST_dataset}}} 
      & 1, 1 & 96    & 1 & 22 & 8.12 & 186.21 \\
    ~ & 1, 2 & 720   & 1 & 37 & 38.45 & 1495.43 \\
    ~ & 1, 3 & 3360  & 1 & 51 & 123.53 & 6235.05 \\
    ~ & 1, 4 & 10920 & 3 & 63 & 323.82 & 7260.88 \\
    ~ & 1, 5 & 26208 & 2 & 76 & 599.03 & 26699.64 \\
    ~ & 1, 6 & 48048 & 1 & 86 & 1052.65 & 101480.30 \\
    \cline{2-7}
    ~ & 2, 2 & 1800   & 1 & 39 & 79.19 & 3503.58 \\
    ~ & 2, 3 & 8400   & 1 & 53 & 292.28 & 17008.22 \\
    ~ & 2, 4 & 27300  & 1 & 66 & 750.19 & 58058.85 \\
    ~ & 2, 5 & 65520  & 1 & 78 & 1508.16 & 131966.71  \\
    ~ & 2, 6 & 120120 & 7 & 90 & 2488.02 & 34123.24 \\
    \hline
    \multirow{6}*{\rotatebox{90}{EMNIST \cite{EMNIST}}} 
      & 1, 1 & 200   & 1 & 56 & 7.38 & 400.06 \\
    ~ & 1, 2 & 1900  & 1 & 94 & 41.40 & 3801.20 \\
    ~ & 1, 3 & 11400 & 1 & 133 & 165.04 & 24079.23 \\
    ~ & 1, 4 & 48450 & 3 & 169 & 566.97 & 31171.86 \\
    \cline{2-7}
    ~ & 2, 2 & 8550   & 1 & 105 & 153.07 & 16653.40 \\
    ~ & 2, 3 & 51300  & 1 & 142 & 683.22 & 99558.56 \\
    ~ & 2, 4 & 218025 & 6 & 176 & 2519.70 & 69996.76 \\
    \hline
  \end{tabular}
  \vspace{-2mm}
\end{table}

\vspace{1mm}\noindent\textbf{Pruning-based Selection.} We run the pruning-based selection algorithms for the datasets with ground truth selection performance and show their performance in Table \ref{tab:selection_pruning}. For each $\{n_{11}, n_{21}\}$ pair, we run four pruning algorithms and report the quality score $s$ of each selected core neuron set, the rank of each selection (based on the ground truth in Table \ref{tab:all_combinations_of_lenet5}), the efficiency ($T$, in seconds) of the selection process, and the $A_s$ and $A_r$ of the protected DNN. For instance, when we designate one neuron from each of the first two convolution layers to be FHE-encrypted and use PFEC pruning to select these neurons, the network achieves 98.50\% $A_s$ (moderate performance) and 98.46\% $A_r$ (bad protection). The corresponding $s$ score of -0.20 ranked 48 out of all 96 possible selections. 

As shown in the table, the two static pruning methods, PFEC and FPGM, are extremely fast, while the dynamic pruning methods, HRank and GFS, are slow. The core neuron sets selected by all four algorithms demonstrate solid $A_s$ in most cases, while $A_r$ appears too high to provide enough protection. Since the dynamic pruning algorithms are significantly slower while they do not provide better selections, we do not consider them in the rest of the paper.

\begin{table*}[t]
  \centering
  \caption{Performance of pruning-based selection algorithms for LeNet-5.}
  \label{tab:selection_pruning}\vspace{-3mm}
  \setlength\tabcolsep{3.2pt}
  \begin{tabular}{c|c|c|c|c|c|c|c|c|c|c|c|c|c|c|c|c|c}
    \hline
    \multirow{2}*{} & $n_{1,1}$, & \multicolumn{4}{c}{PFEC \cite{PFEC}} \vline & \multicolumn{4}{c}{FPGM \cite{FPGM}} \vline & \multicolumn{4}{c}{HRank \cite{hrank}} \vline & \multicolumn{4}{c}{GFS \cite{GFS}} \\
    \cline{3-18}
    ~ & $n_{2,1}$ & \# & $s$ & T & $A_s\ \ \ \ A_r$ & \# & $s$ & T & $A_s\ \ \ \ A_r$ & \# & $s$ & T & $A_s\ \ \ \ A_r$ & \# & $s$ & T & $A_s\ \ \ \ A_r$ \\
    \hline
    \multirow{11}*{\rotatebox{90}{ MNIST \cite{MNIST_dataset}}} & 1,1 
        & 48 & -0.20 & 3 & 98.50 98.46 
        & 44 & -0.19 & 2 & 98.48 98.18 
        & 16 & -0.03 & 384 & 99.24 99.16 
        & 50 & -0.20 & 76 & 98.34 97.70 \\
    ~ & 1,2 
        & 229 & -0.06 & 3 & 99.00 98.32 
        & 237 & -0.06 & 2 & 98.88 97.72 
        & 172 & -0.01 & 384 & 99.26 99.00 
        & 526 & -0.21 & 128 & 98.28 97.48 \\
    ~ & 1,3 
        & 1531 & -0.03 & 3 & 99.04 98.02 
        & 1273 & 0.00 & 2 & 99.02 97.36 
        & 1221 & 0.00 & 385 & 99.28 98.82 
        & 2670 & -0.17 & 176 & 98.22 96.40 \\
    ~ & 1,4 
        & 6541 & 0.00 & 3 & 99.10 97.74 
        & 1936 & 0.16 & 2 & 99.12 94.40 
        & 6066 & 0.02 & 386 & 99.28 98.54 
        & 8239 & -0.06 & 220 & 98.54 95.80 \\
    ~ & 1,5 
        & 15858 & 0.08 & 3 & 99.08 96.04 
        & 3301 & 0.28 & 2 & 99.26 92.22 
        & 14714 & 0.09 & 382 & 99.26 96.86 
        & 23347 & -0.06 & 260 & 98.56 96.00 \\
    ~ & 1,6 
        & 27933 & 0.17 & 3 & 99.18 94.54 
        & 3146 & 0.37 & 2 & 99.24 88.14 
        & 29319 & 0.16 & 387 & 99.26 95.36 
        & 28929 & 0.16 & 295 & 98.50 89.14 \\
    \cline{2-18}
    ~ & 2,2 
        & 461 & -0.13 & 3 & 98.04 94.64 
        & 313 & -0.08 & 2 & 98.70 97.10 
        & 1127 & -0.26 & 383 & 98.04 97.44 
        & 1247 & -0.29 & 147 & 95.74 91.98 \\
    ~ & 2,3 
        & 2189 & -0.03 & 3 & 98.16 93.02 
        & 1607 & -0.01 & 2 & 98.90 96.76 
        & 5854 & -0.19 & 384 & 98.26 97.00 
        & 5224 & -0.16 & 191 & 95.68 88.12 \\
    ~ & 2,4 
        & 8576 & 0.02 & 3 & 98.26 92.34 
        & 1937 & 0.14 & 2 & 98.90 93.36 
        & 17152 & -0.07 & 384 & 98.52 95.80 
        & 21766 & -0.14 & 236 & 95.60 87.40 \\
    ~ & 2,5 
        & 11840 & 0.15 & 3 & 98.44 89.14 
        & 3934 & 0.23 & 2 & 98.96 90.92
        & 16661 & 0.11 & 383 & 98.74 92.84 
        & 58709 & -0.12 & 278 & 95.84 87.02 \\
    ~ & 2,6 
        & 27822 & 0.20 & 3 & 98.46 86.54 
        & 3966 & 0.33 & 2 & 98.94 86.10
        & 20026 & 0.24 & 384 & 99.04 91.62 
        & 92539 & 0.01 & 312 & 95.82 75.02 \\
    \hline
    \multirow{7}*{\rotatebox{90}{EMNIST \cite{EMNIST}}} &1,1 
        & 121 & -0.28 & 3 & 91.71 91.35 
        & 64 & -0.15 & 2 & 92.26 91.46 
        & 19 & 0.03 & 1105 & 92.78 90.75 
        & 25 & 0.20 & 213 & 93.02 92.35 \\
    ~ & 1,2 
        & 1071 & -0.13 & 3 & 91.98 89.59 
        & 938 & -0.11 & 2 & 92.27 90.51 
        & 434 & 0.03 & 1111 & 92.76 90.48 
        & 253 & 0.10 & 347 & 93.00 90.58 \\
    ~ & 1,3 
        & 4160 & 0.04 & 3 & 92.17 86.49 
        & 4963 & 0.01 & 2 & 92.36 88.62 
        & 3290 & 0.09 & 1101 & 92.80 89.50 
        & 597 & 0.28 & 479 & 92.97 85.64 \\
    ~ & 1,4 
        & 5885 & 0.30 & 3 & 92.35 74.88 
        & 8458 & 0.25 & 2 & 92.43 80.90 
        & 17987 & 0.13 & 1110 & 92.83 88.86 
        & 3919 & 0.34 & 601 & 93.02 84.32 \\
    \cline{2-18}
    ~ & 2,2 
        & 5408 & -0.21 & 3 & 91.46 88.88 
        & 6540 & -0.26 & 2 & 91.27 89.35 
        & 2179 & -0.08 & 1108 & 92.22 89.62 
        & 5621 & -0.22 & 408 & 85.30 78.94 \\
    ~ & 2,3 
        & 7320 & 0.01 & 3 & 91.65 84.28 
        & 41379 & -0.12 & 2 & 91.48 86.93 
        & 24419 & -0.06 & 1130 & 92.16 88.85 
        & 16559 & -0.03 & 537 & 86.78 68.88 \\
    ~ & 2,4 
        & 4823 & 0.27 & 3 & 92.19 73.17 
        & 16401 & 0.16 & 2 & 91.80 77.82 
        & 118606 & -0.02 & 1118 & 92.17 88.04 
        & 122706 & -0.02 & 656 & 86.38 67.17  \\
    \hline
  \end{tabular}
\end{table*}

\vspace{1mm}\noindent\textbf{Pruning+Greedy Selection.} Finally, we evaluate the performance of the Pruning+Greedy approach (Algorithm \ref{alg:pruning_greedy}) using PFEC \cite{PFEC} and FPGM \cite{FPGM} pruning methods and compare the performance with the greedy approach. The performance on the smaller MNIST and EMNIST datasets is shown in Table~\ref{tab:selection_mnist}. From the results, we have the following observations:

\noindent  $\bullet$  \textbf{1.} The accuracy of authorized DNN evaluation, $A_s$, remains high (very close to $A_o$) for almost all $\{n_{11}, n_{21}\}$ pairs.  The usability goal defined in Section \ref{subsec:objectives} is always satisfied. 

\noindent  $\bullet$  \textbf{2.} The accuracy of unauthorized DNN evaluation, $A_r$, could be high when only a small number of neurons are encrypted. With the increase of $n_{11}$ and $n_{21}$ (i.e., more neurons in the first two convolution layers are protected), $A_r$ decreases accordingly. In general, the unauthorized user only gets approximately 70\% accuracy when five neurons from the first two convolution layers (30 neurons in total) of the LeNet-5 model for EMNIST are encrypted. The security goal defined in Section \ref{subsec:objectives} is also satisfied.

\noindent  $\bullet$  \textbf{3.} The Pruning+Greedy mechanism saves core neuron selection time ($T$) significantly when $n_{i1} << n_i$, i.e., when a relatively small number of neurons in a layer will be encrypted. However, the improved efficiency comes at the cost of selection performance. Meanwhile, there does not exist any obvious differences in the performance of PFEC and FPGM. 

\begin{table}[tbp]
  \centering
  \vspace{-2mm}
  \caption{Performance of the Pruning+Greedy approach for MNIST and EMNIST.}
  \label{tab:selection_mnist}\label{tab:selection_emnist}\vspace{-3mm}
  \setlength\tabcolsep{1.6pt}
  \begin{tabular}{c|c|c|c|c|c|c|c|c|c}
    \hline
    \multirow{3}*{\makecell{$n_{1,1}$,\\$n_{2,1}$}} & \multicolumn{3}{c}{\multirow{2}*{Greedy}} \vline & \multicolumn{6}{c}{Pruning+Greedy} \\
    \cline{5-10}
    ~  & \multicolumn{3}{c}{} \vline & \multicolumn{3}{c}{PFEC \cite{PFEC}} \vline & \multicolumn{3}{c}{FPGM \cite{FPGM}} \\
    \cline{2-10}
    ~ & \# & T & $A_s\ \ \ A_r$ & \# & T & $A_s\ \ \ A_r$ & \# & T & $A_s\ \ \ A_r$ \\
    \hline\hline\multicolumn{10}{c}{MNIST \cite{MNIST_dataset}} \\\hline 
    1,1 
        & 1 & 22 & 99.30 95.66 
        & 9 & 5 & 99.26 98.84 
        & 5 & 6 & 99.28 98.68 \\
    1,2 
        & 1 & 37 & 99.30 89.30 
        & 84 & 10 & 99.32 98.54 
        & 48 & 11 & 99.34 97.92 \\
    1,3 
        & 1 & 51 & 99.28 86.04 
        & 437 & 18 & 99.36 97.34 
        & 322 & 19 & 99.34 96.46 \\
    1,4 
        & 3 & 63 & 99.22 81.12 
        & 1430 & 29 & 99.32 94.98 
        & 1430 & 30 & 99.32 94.98 \\
    1,5 
        & 2 & 76 & 99.26 77.66 
        & 824 & 43 & 99.24 88.88 
        & 824 & 44 & 99.24 88.88 \\
    1,6 
        & 1 & 86 & 99.34 76.64 
        & 2296 & 60 & 99.22 87.02 
        & 2296 & 61 & 99.22 87.02 \\
    \hline
    2,2 
        & 1 & 39 & 98.66 87.32 
        & 282 & 12 & 98.50 95.76 
        & 262 & 14 & 98.62 96.28 \\
    2,3 
        & 1 & 53 & 98.86 85.36 
        & 754 & 20 & 98.60 94.00 
        & 417 & 22 & 98.62 93.18 \\
    2,4 
        & 1 & 66 & 99.06 82.24 
        & 906 & 31 & 98.92 92.04 
        & 833 & 32 & 98.78 90.62 \\
    2,5 
        & 1 & 78 & 99.18 82.32 
        & 147 & 46 & 99.12 87.44 
        & 147 & 46 & 99.12 87.44 \\
    2,6 
        & 7 & 90 & 99.22 80.82 
        & 59 & 63 & 99.12 80.80 
        & 59 & 63 & 99.12 80.80 \\
    \hline\hline\multicolumn{10}{c}{EMNIST \cite{EMNIST}} \\\hline 
    1,1 
        & 1 & 56 & 93.02 89.21 
        & 16 & 9 & 93.04 92.04 
        & 20 & 11 & 93.02 92.21 \\
    1,2 
        & 1 & 94 & 93.02 81.37 
        & 3 & 20 & 93.00 83.57 
        & 15 & 21 & 93.03 86.23 \\
    1,3 
        & 1 & 133 & 93.03 71.62 
        & 7 & 37 & 93.04 77.58 
        & 3 & 38 & 93.03 73.92 \\
    1,4 
        & 3 & 169 & 93.06 66.23 
        & 64 & 59 & 93.04 73.64 
        & 5 & 60 & 93.05 67.06 \\
    \hline
    2,2 
        & 1 & 105 & 92.86 80.91 
        & 6 & 25 & 92.79 83.35 
        & 11 & 26 & 92.91 85.64 \\
    2,3 
        & 1 & 142 & 92.85 70.85 
        & 36 & 42 & 92.76 77.12 
        & 2 & 43 & 92.88 73.33 \\
    2,4 
        & 6 & 176 & 92.88 63.81 
        & 343 & 65 & 92.77 72.99 
        & 6 & 65 & 92.88 63.81 \\ 
    \hline
  \end{tabular}
  \vspace{-5mm}
\end{table}

Finally, we report the performance of greedy and pruning+greedy approaches on the three complex datasets, GTSRB, CIFAR10, and Tiny-ImageNet. We evaluate these algorithms when 15\% to 75\% of the neurons in the first two convolution layers are selected to be encrypted. Since we do not have the ground truth (ranking) for all possible core neuron sets, we only report $A_s$, $A_r$, and $T$ in Table \ref{tab:selection_gtsrb}. As shown, even if we select 25\% neurons from the first two convolutional layers, all the models' usability and security goals are clearly satisfied. Since we use $p=2$ for pruning+greedy, when 50\% or more of the neurons are to be encrypted, there is no need for pruning. The pruning+greedy approach provides slightly worse selections while being more efficient. In summary, the pruning+greedy approach balances selection efficiency and performance compared to the greedy algorithm. 
However, since neuron selection is a one-time process, the model owner may want to employ the greedy approach to obtain better performance while accepting the overhead. 

\begin{table}[tbp]
  \centering\vspace{-3mm}
  \caption{Performance of the Pruning+Greedy approach for GTSRB, CIFAR10, and Tiny-ImageNet.}
  \label{tab:selection_gtsrb}\label{tab:selection_cifar10}\vspace{-3mm}
  \setlength\tabcolsep{2.5pt}
  \begin{tabular}{c|c|c|c|c|c|c}
    \hline
    \multirow{3}*{\makecell{$n_{i,1}$ \\ (\%)}} & \multicolumn{2}{c}{\multirow{2}*{Greedy}} \vline & \multicolumn{4}{c}{Pruning+Greedy} \\
    \cline{4-7}
    ~  & \multicolumn{2}{c}{} \vline & \multicolumn{2}{c}{PFEC \cite{PFEC}} \vline & \multicolumn{2}{c}{FPGM \cite{FPGM}} \\
    \cline{2-7}
    ~ & T & $A_s\ \ \ A_r$ & T & $A_s\ \ \ A_r$ & T & $A_s\ \ \ A_r$ \\
    \hline\hline\multicolumn{7}{c}{GTSRB \cite{GTSRB}} \\\hline 
    15 & 35703 & 94.05 89.33 & 8656 & 93.90 86.94 & 8798 & 93.94 86.33 \\
    25 & 56082 & 94.17 79.90 & 24665 & 94.06 79.16 & 24560 & 94.00 77.83\\
    50 & 104842 & 94.27 68.95 & - & - & - & - \\
    75 & 121636 & 94.25 37.96 & - & - & - & - \\
    \hline\hline\multicolumn{7}{c}{CIFAR10 \cite{CIFAR10_dataset}} \\\hline 
    15 & 2924 & 90.90 64.20 & 679 & 89.82 53.36 & 686 & 90.14 53.20 \\
    25 & 4899 & 91.18 60.34 & 2121 & 90.32 66.72 & 2144 & 90.32 66.72 \\
    50 & 8336 & 90.78 55.90 & - & - & - & - \\
    75 & 10479 & 90.52 24.90 & - & - & - & - \\   
    \hline\hline
    
    \multicolumn{7}{c}{\textcolor{black}{Tiny-ImageNet \cite{tinyimagenet}}} \\\hline 
    \textcolor{black}{15} & \textcolor{black}{3954} & \textcolor{black}{72.10 44.84} & \textcolor{black}{911} & \textcolor{black}{71.10 39.06} & \textcolor{black}{920} & \textcolor{black}{70.96 37.94} \\
    \textcolor{black}{25} & \textcolor{black}{6626} & \textcolor{black}{72.16 43.18} & \textcolor{black}{2842} & \textcolor{black}{71.48 39.30} & \textcolor{black}{2889} & \textcolor{black}{71.18 42.52} \\
    \textcolor{black}{50} & \textcolor{black}{11388} & \textcolor{black}{72.02 45.66} & - & - & - & - \\
    \textcolor{black}{75} & \textcolor{black}{14289} & \textcolor{black}{72.04 26.76} & - & - & - & - \\   
    \hline
  \end{tabular}
  \begin{minipage}{8.5cm}
  \vspace{0.1cm}
  \small  Notes: \textbf{1.} $A_o$ of GTSRB and CIFAR10 are 93.24\% and 90.22\%, respectively.
  \end{minipage} \vspace{-2mm}
\end{table}

\subsection{Computation for DNN Evaluation}\label{subsec:computationexp}

\begin{figure}[t]
    \vspace{-4mm}\centerline{\includegraphics[width=0.96\columnwidth]{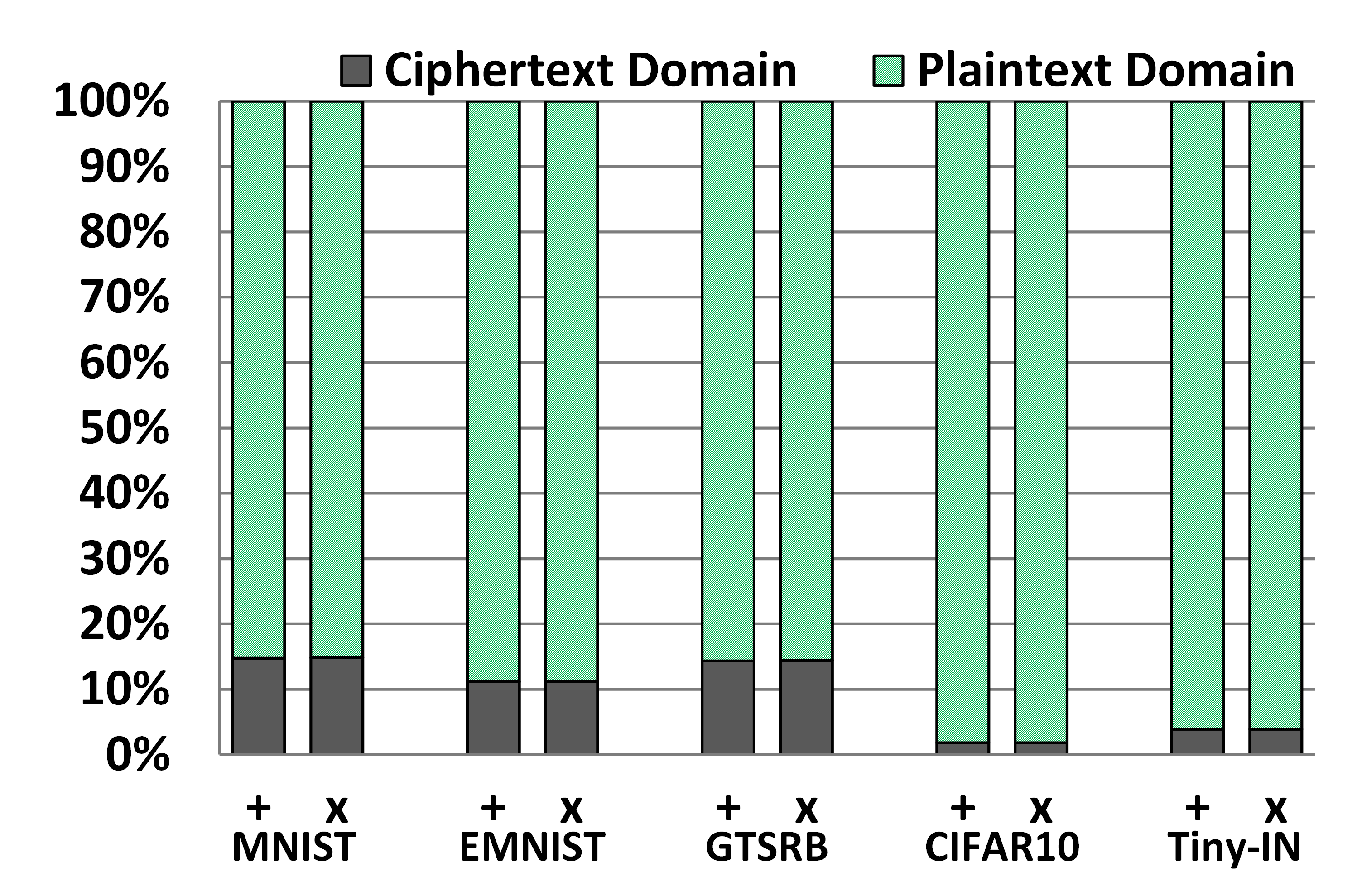}} 
    \vspace{-4mm}
    \caption{Distribution of addition (+) and multiplication ($\times$) operations in the ciphertext and plaintext domains.}
    \label{fig:proportion}\vspace{-3mm}
\end{figure}

The essential advantage of employing partial DNN encryption in \privdnn~is to reduce the heavy-lifting ciphertext operations. Instead of encrypting the full network and performing the entire DNN evaluation process using FHE in ciphertext, \privdnn~only protects the \textit{core} of the target DNN, i.e., the neurons that are essential to the performance. Only a subset of the neurons are encrypted, and the corresponding operations are performed in the ciphertext domain, while the majority of the neurons and the corresponding operations are kept intact in the plaintext domain. 

We define the DNN encryption ratio $ER = \frac{N_e}{N}$, i.e., the proportion of core neurons out of all neurons in the DNN. We also define the encryption ratio of addition and multiplication operations as $ER_{+}$ and $ER_{*}$, denoting the proportion of ciphertext addition and multiplication operations out of all operations, respectively. Compared to FHE operations, the computation costs of plaintext operations are negligible. Meanwhile, the multiplication operations (usually followed by a rescale operation), cost more than ten times time than the addition operations. Therefore, the inference time of \privdnn~is almost linear to $ER_{*}$, which is mostly determined by $ER$.

In Figure \ref{fig:proportion}, we demonstrate the distribution of operations (addition and multiplication) in the ciphertext and plaintext domains. For MNIST and EMNIST, we use $\{n_{1,1}=1, n_{2,1}=3\}$. For the larger models, we encrypt 50\% of the neurons in the first two convolution layers. As shown in Figure \ref{fig:proportion}, \privdnn~only places a very small portion of the operations in the ciphertext domain, while the majority of the operations remain in plaintext. Therefore, compared with the full DNN encryption approaches in the literature, e.g., \cite{cryptonets}, \privdnn~saves 85\% to 98\% of the cipher computation. Besides, the traditional full DNN encryption approach requires a higher HE noise budget, slowing down all cipher operations. 
For example, when we implement LeNet-5 on the MNIST dataset, the \privdnn~inference for a batch of 5,000 testing images costs 190 seconds with the selection of $\{n_{1,1}=1, n_{2,1}=4\}$. In comparison, the full DNN encryption approach takes 5,647 seconds. That is, \privdnn~achieves a 29.7x speedup with $ER_{*}=18.75\%$. 

Besides the efficiency benefits, \privdnn~also saves a significant amount of memory so that the deployment of FHE-based inference for large-scale DNN models becomes practical for commodity desktop computers. The traditional full DNN encryption approach needs a higher noise budget to support ciphertext operations along all the DNN layers, which causes the cipher to take up more memory for each ciphertext value. For example, the ciphertext weights of the first fully connected layer of LeNet-5/MNIST take 151 GB, which is beyond the capacity of most PCs. On the contrary, \privdnn~uses approximately 2.5 GB of memory with $\{n_{1,1}=1, n_{2,1}=3\}$, and 5 GB with $\{n_{1,1}=2, n_{2,1}=6\}$. Therefore, \privdnn~dramatically reduces the memory utilization for FHE-based DNN inference and consequently enables the deployment of secure DNN inference for resource-constrained users such as small businesses.

\vspace{2mm}\noindent\textbf{Summary of Experimental Results.} From the experiments, we conclude that: (1) The greedy algorithm effectively identifies a core neuron set that achieves high accuracy for authorized DNN execution and low accuracy for unauthorized execution. (2) Encrypting approximately 15\% to 25\% of the neurons in the first two convolution layers of a large DNN effectively reduces its accuracy by 20\% for unauthorized users, i.e., the DNN becomes practically unusable for them. (3) With the partial DNN encryption scheme, only a few DNN evaluation operations are conducted in ciphertext. Compared with the full-DNN encryption scheme, \privdnn~saves the cipher operations by 85\% to 98\%, reduces the inference time and memory usage by 97\% on smaller models and much more on larger models.

\section{Security Analysis}\label{sec:securityanalysis}

We discuss the security/privacy guarantees of the proposed \privdnn\ framework. We consider two aspects of privacy: the model owners would like to protect their proprietary deep-learning models, while the data owners would like to protect their testing samples. 

\subsection{Model Privacy}\label{subsec:modelprivacy}

\noindent\textbf{Privacy Expectations}. The downstream user, i.e., the data owner $\mathcal{C}$, should not be able to utilize the partially encrypted DNN to reconstruct a model to achieve comparable accuracy to the original model. We discuss the following aspects: (1) information that is disclosed to the data owner $\mathcal{C}$, (2) the capability of honest data owners, (3) the capability of curious data owners, and (4) the capability of dishonest data owners and the potential defense.

\noindent\textbf{Structure and Parameters.} \privdnn~allows $\mathcal{C}$ to run some inference operations in the plaintext domain to avoid expensive cipher domain operations. Therefore, $\mathcal{S}$ must share the model structure and the non-core parameters in plaintext with $\mathcal{C}$. $\mathcal{S}$ also shared the core parameters in ciphertext with $\mathcal{C}$. 

\noindent\textbf{The Honest Data Owners.} $\mathcal{C}$ receives the core neuron weights in ciphertext. As explained in Section \ref{subsec:threat_model}, the confidentiality of the encrypted weights is guaranteed by the security of the FHE algorithm CKKS. It is proved that the security CKKS relies on the hardness of Ring-LWE, which indicates that if there is an adversary who can break CKKS, it can be used to tackle the RLWE problem. It is believed that there is only a negligible probability of this. In this regard, $\mathcal{C}$ cannot break the encryption to learn the weights, and hence, it could only achieve $A_r$ with the non-core parameters. The core neuron selection algorithms ensure that $A_r$ is always much lower than $A_s$ (accuracy for the authorized users). This is also demonstrated by experiments in Tables \ref{tab:selection_mnist} and \ref{tab:selection_gtsrb}. 

{\color{black}\noindent\textbf{The Curious Data Owners and the Model Recovery Attack.} A curious data owner ($\mathcal{C}$) precisely follows the SMC protocol in model execution, but she still attempts to infer knowledge about the protected model from the information she received and the intermediate results. We define the model recovery attack as follows: the attacker (curious data owner) knows the target model’s architecture and a subset of plaintext weights, while the remaining weights are encrypted. According to the threat model (Section \ref{subsec:threat_model}), the attacker cannot break the encryption, nor does she have the model owner’s proprietary data to train the model. The attacker uses a small number of labeled samples to \textit{retrain} the model in an attempt to recover the unknown weights and restore the model.} In particular, to get the model with comparable accuracy to the original model, the attacker executes the model recovery attack as follows: (1) The attacker removes all the encrypted weights from the protected DNN and reset them. (2) With her local samples, the attacker tunes the entire model or freezes the plaintext parameters and only trains the missing weights. Our experiments prove that the first scenario is always better, so we adapt the first scenario.

To evaluate the attacker's ability, we run the experiments with two settings: [Setting I] Only a smaller amount of neurons are encrypted to provide less protection with high efficiency: $n_{1,1}=1, n_{2,1}=2$ for MNIST and EMNIST, 50\% of the neurons in the first two convolution layers for GTSRB and CIFAR10. [Setting II] More neurons are encrypted for better protection: $n_{1,1}=2, n_{2,1}=4$ for MNIST and EMNIST, 75\% of the neurons in the first two convolution layers for GTSRB and CIFAR10. (still only 2.27\% of all convolution neurons for CIFAR10). To tune the model, we use the same parameters as the original training process except for a one-tenth learning rate for the selected neurons and 64 epochs.

The experimental results for the recovery attacks are presented in Table \ref{tab:recover}, where $A_{rec}$ denotes the accuracy of the recovered models. Results show that DNNs protected by \privdnn~are relatively resilient against recovery attacks. In particular: (1) In most cases, the models recovered from partially encrypted DNNs perform noticeably worse than the protected models. Note that, model accuracy does not linearly increase with training resources/efforts. Training a moderately accurate model from a modest amount of data is relatively easy, while it is exponentially more challenging to improve the accuracy for another 5\% or even 3\% from a (publicly available) modest model. Therefore, a small accuracy drop of the recovered models dramatically reduces their values. (2) In some cases, the recovered models perform worse than models trained from scratch with 1,000 samples ($A_{t}$). As discussed in the threat model, the data owners are not expected to have too many samples for the model recovery attack. Otherwise, they could train the model by themselves. (3) Large models are more valuable to protect in real-world practice. For large models such as VGG16 for CIFAR10, \privdnn~could encrypt a relatively larger portion of the first two convolution layers, making it highly challenging for the data owners to recover the network. Meanwhile, since the networks are deep and complex, most of the computations are still in plaintext, even when the first two convolution layers are mostly encrypted.

\begin{table}[t]
  \centering
  \caption{Recover attack by curious data owners.}
  \label{tab:recover}\vspace{-3mm}
  \setlength\tabcolsep{2.12pt}
  \begin{tabular}{c|c|c|c|c|c|c|c|c|c|c}
    \hline
    \multicolumn{11}{c}{Model recovery from partially encrypted DNN} \\
    \hline
    & \multicolumn{5}{c}{Setting 1} \vline & \multicolumn{5}{c}{Setting 2} \\
    \cline{2-11}
    ~ & \multirow{2}*{$A_s\ \ A_r$} & \multicolumn{4}{c|}{$A_{rec}$}
    & \multirow{2}*{$A_s\ \ A_r$} & \multicolumn{4}{c}{$A_{rec}$} \\\cline{3-6}\cline{8-11}
    ~ & ~ & 100 & 250 & 500 & 1000 & ~ & 100 & 250 & 500 & 1000 \\
    \hline
    M & 99.3 89.3 & 89.1 & 89.2 & 89.5 & 90.5 
        & 99.1 82.2 & 92.8 & 93.0 & 93.3 & 94.4  \\
    E & 93.0 81.4 & 81.5 & 84.9 & 88.7 & 89.8 
        & 92.9 63.8 & 66.1 & 74.9 & 83.9 & 86.1 \\
    G & 94.3 69.0 & 86.1 & 89.7 & 91.8 & 92.7 
        & 94.3 38.0 & 68.7 & 79.7 & 86.1 & 90.4 \\
     C & 90.8 55.9 & 73.4 & 78.9 & 80.6 & 83.6 
         & 90.5 24.9 & 21.8 & 19.7 & 48.2 & 58.2 \\
    \hline
  \end{tabular}
  \setlength\tabcolsep{5.2pt}
  \begin{tabular}{c|c|c|c|c|c|c|c|c|c}
    \hline
    \multicolumn{10}{c}{$A_t$} \\
    \hline
    ~ & 100 & 250 & 500 & 1000 & ~ & 100 & 250 & 500 & 1000 \\
    \hline
    M & 59.8 & 78.3 & 85.9 & 91.8 & E & 29.4 & 44.7 & 58.6 & 68.5 \\
    G & 27.6 & 49.9 & 72.5 & 91.5 & C & 23.5 & 31.2 & 39.5 & 49.9 \\
    \hline
  \end{tabular}
  
  \begin{minipage}{8.45cm}
  \vspace{0.1cm}
  \small  {Notes: \textbf{1.} 100, 250, 500, and 1000: the number of pictures to train the model to recover encrypted weights. \textbf{2.} $A_{rec}$: accuracy of the recovered model. {3.} $A_{t}$: accuracy of training from scratch.}
  \end{minipage}
\end{table}

\begin{figure}[t]
\centerline{\includegraphics[width=0.88\columnwidth]{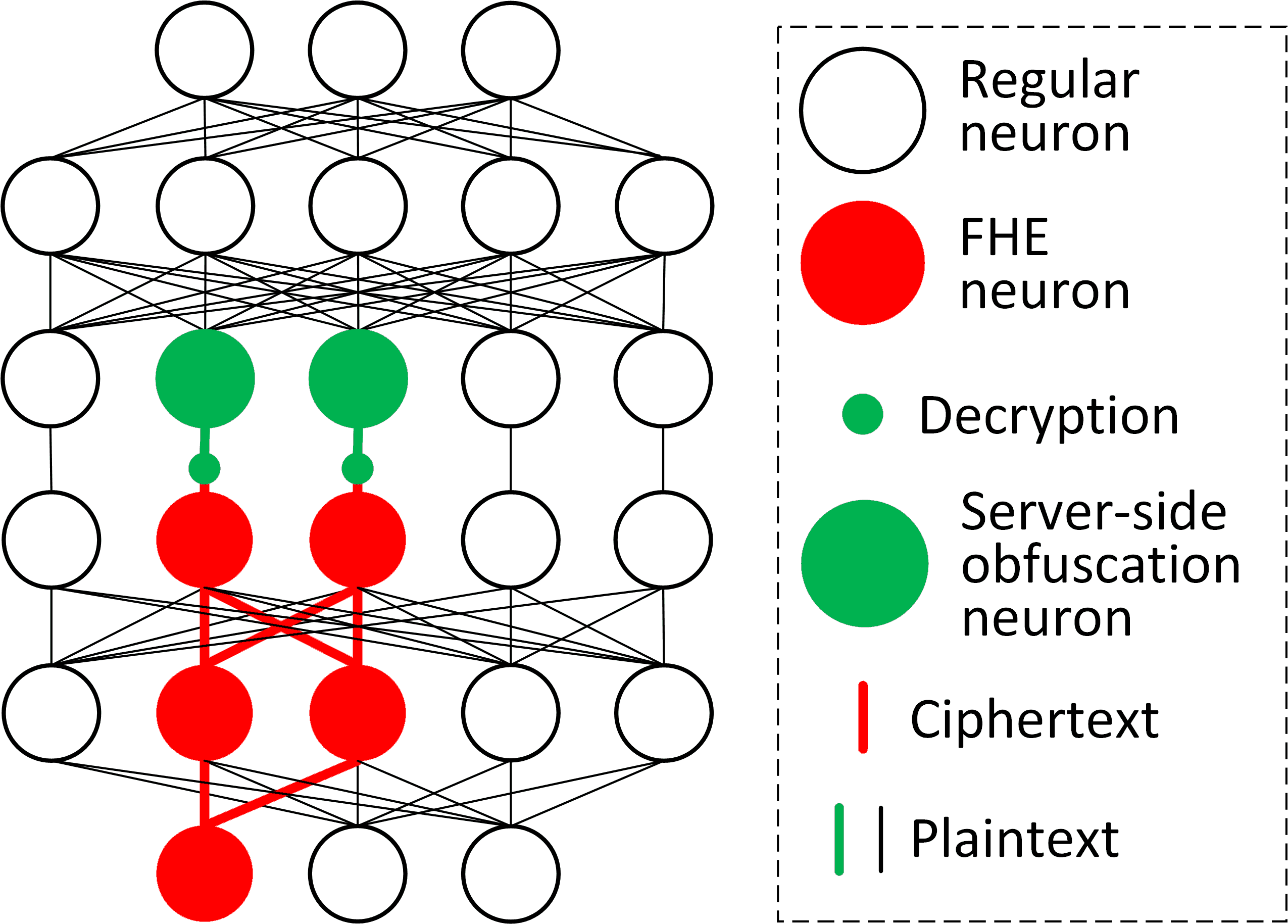}}
    \vspace{-2mm}
    \caption{\privdnn~with a server-side obfuscation layer.} 
    \label{fig:server-obf}\vspace{-3mm}
\end{figure}

\subsection{Data Privacy}

\noindent\textbf{Privacy Expectations.} The model owner, $\mathcal{S}$, should not be able to reconstruct the raw pixels of any testing image or to recover the visual features of any testing image. We discuss the following aspects: (1) information that is (not) disclosed to the model owner, (2) the capability of the honest-but-curious model owners, and (3) the capability of dishonest model owners and the potential defense. 

\vspace{1mm}\noindent\textbf{Raw Testing Samples.} In the design of \privdnn, most inference operations are executed on the data owner ($\mathcal{C}$) side. Therefore, the raw pixel space representations of the testing samples never leave $\mathcal{C}$, and they are considered secure.  

\vspace{1mm}\noindent\textbf{The Curious Model Owners and the Sample Inference Attack.} When $\mathcal{C}$ sends the encrypted output of the core neurons to $\mathcal{S}$ for decryption, $\mathcal{S}$ learns the decrypted values of the intermediate DNN outputs. A curious $\mathcal{S}$ will attempt to utilize the intermediate outputs to infer features of the data samples. This is highly challenging, if not impossible, since (1) she only sees a subset of outputs from the last partially encrypted layer, and (2) $\mathcal{C}$ will follow the protocol to obfuscate the intermediate results before sending them for decryption. In particular, we have adopted two DNN input sample recovery attacks in the literature, split learning \cite{erdougan2022unsplit} and autoencoder \cite{autoencoder}, in an attempt to recover C’s samples from the obfuscated plaintext output of the core neurons. As shown in Figure \ref{fig:data-recovery-unsplit100} and \ref{fig:data-recovery-autoencoder100}, the attacks could not succeed even when 100\% of the Layer 2 output (obfuscated) is known to $\mathcal{S}$.

\subsection{The Dishonest Participants}

\vspace{2mm}\noindent\textbf{The Threat Model Revisited.} We would like to note that the honest-but-curious model is adopted in this explorative project on SMC for deep learning. However, the honest assumption could be too strong for real-world adoption. Here, we attempt to relax this assumption to explore the simple attacks from malicious parties and propose corresponding controls. We acknowledge that more research efforts need to be devoted to further investigating the complex attacks and defenses to fully relax the honest assumption.

\vspace{1mm}
\noindent\textbf{The Dishonest Data Owners and the Weight-Stealing Attack.}
A malicious $\mathcal{C}$ may break the honest-but-curious assumption to launch a \textit{weight-stealing attack}: it sends the encrypted and obfuscated model parameters to $\mathcal{S}$, who decrypts and returns the weighs. Hence, $\mathcal{C}$ will get the whole plaintext model. 

To defend against the attack, we infuse a server-side obfuscation layer after the last partially encrypted layer, as shown in Figure \ref{fig:server-obf}. The obfuscation layer is a convolutional layer (5$\times$5 kernels) with only one input and output channel. The weights for the neurons directly succeeding the encrypted core neurons are kept with $\mathcal{S}$, while the weights for the other neurons in this layer are shared in plaintext with $\mathcal{C}$. This layer is used to obfuscate the decryption results before returning them to $\mathcal{C}$, that is, $\mathcal{S}$ decrypts the intermediate results, performs convolution operations with the server-side obfuscation neurons, and returns the obfuscated results to $\mathcal{C}$. Since $\mathcal{C}$ does not have the parameters of the obfuscation neurons, even if she sends the encrypted model weights to $\mathcal{S}$, $\mathcal{C}$ cannot recover such weights after server-side obfuscation. More importantly, the obfuscation layer is polymorphic, i.e., when we freeze the entire network, reset and retrain the obfuscation neurons (the green neurons in Figure \ref{fig:server-obf}), we obtain a different set of the obfuscation parameters in each retrain. The polymorphism feature makes it impossible for a malicious $\mathcal{C}$ to reverse engineer the parameters of the obfuscation neurons. Note that the computation of the obfuscation layer is negligible since it is only a convolution layer in plaintext. In our experiments, this polymorphic obfuscation layer only decreases the model's accuracy by no more than 0.1\%. Server-side obfuscation is also compatible with client-side obfuscation. In client-side obfuscation, $\mathcal{C}$ multiplies all the encrypted values from the same channel by a random value $\sigma$ before sending them to $\mathcal{S}$. Since the operations in the server-side obfuscation (convolution) layer are all linear, $\mathcal{C}$ could de-obfuscate the results by multiplying by $1/\sigma$. 

\vspace{1mm}\noindent\textbf{Dishonest Model Owner.} A malicious $\mathcal{S}$ may use engineered core weights (e.g., using all 1s and 0s) to obtain (partial) plaintext input that was sent into the core neurons. This is prevented by data obfuscation. Besides, with the malicious core weights, $A_s$ will dramatically decrease, and $\mathcal{C}$ easily detects the malicious model.

\subsection{The Performance-Security Trade-off}\label{subsec:tradeoff}

\begin{figure*}[t]
    \includegraphics[width=\textwidth]{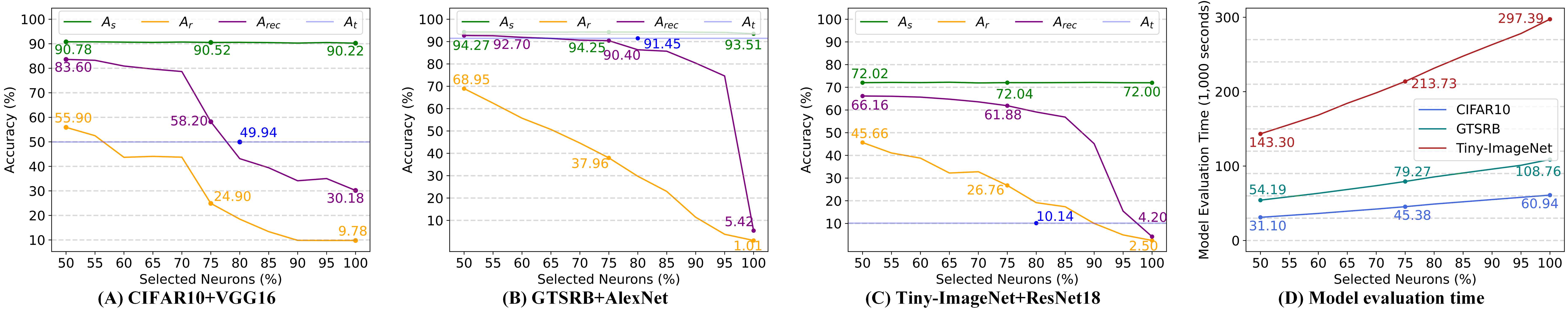}
  \vspace{-7mm}
  \caption{The performance-security trade-off: the distribution of $A_s$, $A_r$ when 50\% to 100\% neurons are encrypted. The corresponding $A_{rec}$ and $A_t$ when 1,000 samples are used for model recovery attacks or train the model for (A) CIFAR10+VGG16, (B) GTSRB+AlexNet, and (C) Tiny-ImageNet+ResNet18. (D): Model evaluation time ($T$) for a batch of up to 8192 testing samples.}\label{fig:recovery-attack-all}
\vspace{-1mm}
\end{figure*}

In \privdnn, $\mathcal{S}$ has full control of the selection of the core neurons. In practice, $\mathcal{S}$ identifies $n_{i,1}$, which is the number of core neurons to be selected from layer $i$. \privdnn~then selects the (near) optimal set of core neurons that achieve the best performance for authorized users ($A_s$) and the least performance for unauthorized users ($A_r$). On the other hand, the choice of parameters $n_{i,1}$ implies a trade-off between performance and security. If $\mathcal{S}$ chooses to encrypt fewer core neurons (smaller $n_{i,1}$), it saves computation at $\mathcal{C}$ but provides weaker protection, i.e., a higher $A_r$ and potentially more vulnerable against the model recovery attack (higher $A_{rec}$).  On the contrary, if $\mathcal{S}$ encrypts more core neurons, it will require higher computation at $\mathcal{C}$ but provide better protection for $\mathcal{S}$'s model. 

{To demonstrate the performance-security trade-off, we evaluate \privdnn~on three datasets: CIFAR10+VGG16, GTSRB+AlexNet, and Tiny-ImageNet+ResNet18. We protect 50\% to 100\% of the neurons from the first two convolution layers. The results are shown in Figure \ref{fig:recovery-attack-all}. We can observe the following: (1) we achieve consistently high $A_s$ (accuracy for authorized model evaluation), which is always very close to the original model accuracy ($A_o$). (2) $A_r$ (accuracy for unauthorized evaluation) decreases with the increased number of encrypted neurons, i.e., from 55.90\% to 9.78\% for CIFAR10, from 68.95\% to 1.01\% for GTSRB, and from 45.66\% to 2.50\% for Tiny-ImageNet. (3) Completely retraining the model with 1000 samples achieves subpar accuracy ($A_t$): 49.94\% for CIFAR, 91.45\% for GTSRB, and 10.14\% for Tiny-ImageNet. (4) The effectiveness of the model recovery attack also decreases, especially when $\sim$100\% of the first two layer neurons are protected. When 50\%, 75\%, and 100\% of the first two layers are encrypted, the accuracy of the model recovered with 1,000 samples ($A_{rec}$) decreases from 83.60\%, 58.20\% to 30.18\% for CIFAR-10, from 92.70\%, 90.40\%, to 5.42\% for GTSRB, from 66.16\%, 61.88\% to 4.20\% for Tiny-ImageNet. (5) $A_{rec}$ almost always stays in between $A_s$ and $A_r$, i.e., the model recovery attack could improve the performance of the broken models by tuning them with 1,000 samples. However, they cannot reach the accuracy of the original model, and they fail badly when a relatively larger proportion of the first two layer neurons are protected. (6) The model recovery attack appears more effective on GTSRB. This is explained by the fact that GTSRB has highly similar images in each class, so a classifier could easily achieve high accuracy with only a small number of training samples. As shown in Figure \ref{fig:recovery-attack-all} (B), $A_t$ is very high (approximately 3\% below $A_o$), and $A_{rec}$ stays lower than $A_t$, i.e., with the same number of labeled samples, it is easier to train a model from scratch than to recover a protected model. }

Last, the DNN inference time increases linearly with the number of core neurons, which is consistent with the discussions in Section \ref{subsec:computationexp}. Note that invoking an FHE-based full-model encryption scheme for a large model like VGG16 is extremely expensive. Hence, the overhead for \privdnn~ is still considered very acceptable. 

To balance the efficiency (for $\mathcal{C}$) and model security (for $\mathcal{S}$), the model owner could invoke \privdnn~to evaluate the model with different $n_{i,1}$ settings and empirically pick the one that satisfies her privacy goals while requires reasonable computation for $\mathcal{C}$. Finally, we acknowledge that more effective model recovery attacks may emerge in the future as part of the cat-and-mouse nature of attacks and defenses in cybersecurity. A possible mitigation is to further increase the proportion of the protected neurons. For large models, such as VGG16 and ResNet18, the first two convolution layers contain less than 5\% of all model weights. Encrypting all of them would significantly reduce $A_{rec}$ while still maintaining a low DNN encryption ratio ($ER$), i.e., maintaining high efficiency.

\section{Related Works}\label{sec:relworks}

\noindent\textbf{Privacy-preserving DNN Evaluation with Data Encryption.} Machine learning applications usually involve two steps: training the model from labeled data and inference for unlabeled samples using the trained model. Most of the existing secure ML works focus on protecting the data, i.e., the data owner wants to evaluate the testing samples while not leaking the data to the untrusted model owner. Hence, the data owner encrypts the testing samples using FHE before sending them to the model owner.

CryptoNets \cite{cryptonets} is the first work to apply FHE to secure DNN evaluation. It takes 250 seconds to evaluate a batch of encrypted MNIST images. Later works attempted to improve the efficiency of secure DNN evaluation and support more complex models with higher accuracy. LoLa \cite{low_latency} encrypts entire layers and changes data representations throughout the computation to get an 11.2$\times$ speedup on CIFAR10. Faster CryptoNets \cite{immitate2} improved performance over CryptoNets by leveraging sparsity properties. CHET \cite{CHET} provided an optimizing compiler for FHE DNN inference to offer a high-level user framework to automate parameter tuning. Recently, more efforts have been devoted to improving the activation functions. CryptoDL \cite{cryptodl} explored more common activations besides the square function. FHE–DiNN \cite{precompute_discrete} use a precomputed table to process the cipher activation whose complexity is strictly linear in the depth of the network and whose parameters can be set beforehand. However, it can only be applied to integer ciphers, which means its accuracy is still limited. More works try to use low-degree polynomials to approximate the non-linear activation functions. \cite{immitate1} combine CryptoNets and batch normalization principle. It normalizes the convolutional output before activation and uses different degrees of polynomials to imitate ReLU. \cite{immitate3} examines the contribution of different properties, such as differentiable, continuous, monotonic, etc., to guide the selection of polynomial functions. \cite{sign_function} selects optimal composite polynomials for the sign function and uses the sign function to implement ReLU.

\noindent\textbf{Privacy-preserving DNN Evaluation with Model Encryption.}  Compared with data protection, fewer works focus on model protection. \cite{first_protect_model} first studied the protection of the private DNN models, while \cite{first_protect_cnn_model} focused on the protection of CNNs. They adopt the same data protection schemes for DNN model protection, i.e., they encrypt all the parameters in the model and feed the encrypted model with plaintext or ciphertext data to perform DNN evaluation entirely in ciphertext. Unlike these approaches, \privdnn~is the first to observe and utilize a key feature in DNN model protection that is unnecessary to protect all the parameters in a large DNN. Therefore, \privdnn~is able to achieve a significant boost in the efficiency of privacy-preserving DNN evaluation by only encrypting a subset of essential neurons of the entire DL model.

{\noindent\textbf{DNN Model Recovery Attacks.} Model recovery (stealing or extraction) attacks aim to obtain model hyperparameters, architecture, or trained weights, \cite{model_recovery1,yan2020cache,recovery_attack_optim_1}. The weight-stealing attacks are the most similar to our model recovery attacks. They can be grouped into two categories: (1) stealing exact properties and (2) stealing approximate behavior \cite{model_recovery1}. \cite{first_exact_stealing} proposes the first exact model weight stealing attack on a binary classifier. \cite{copycat} extended the idea and presented an equation-solving attack for support vector regression. 
In outsourced DNN execution, weights could be stolen from the computing platforms or the communication channel \cite{hu2020deepsniffer,yan2020cache,hong2018security}. Such attacks are not applicable in \privdnn, since the cryptography primitives are assumed to be unbreakable to the attacker. 
Meanwhile, in query-based approximate model stealing, the attacker queries the target model with input images to learn the predictions and then trains a local model with learned image-prediction pairs \cite{model_recovery3}. 
Optimization techniques, such as active learning, reinforcement learning, and evolutionary algorithms, are employed to save query budgets \cite{recovery_attack_optim_1, recovery_attack_optim_2, recovery_attack_optim_3}. However, those attacks still need significantly more labeled samples than our model recovery attack. For example, \cite{model_recovery2} stole a ResNet18 model (original accuracy is 78.52\%) and achieved 72.83\% clone accuracy with the entire CIFAR10 dataset (50K samples) as a proxy, or 43.56\% with synthetic fake data. \cite{copycat} stole a VGG16 model and achieved 93.7 - 98.6\% clone accuracy of the original model accuracy with over 3M natural pictures. \cite{recovery_attack_prada} stole a GTSRB\cite{GTSRB} classifying model and achieved 97.9\% clone accuracy of the original model accuracy with 102K synthetic samples.}

\section{Conclusion}\label{sec:conclusion}

We present \privdnn, a practical framework to protect DNN models in privacy-preserving DNN evaluation. With the novel partial DNN encryption scheme, the authorized DNN evaluation accuracy remains very close to the original DNN accuracy, while the unauthorized users get significantly decreased accuracy. We design and implement three algorithms to identify the core neurons from DNN models for effective protection. As shown in extensive experiments on five popular benchmarking datasets and DNN models, \privdnn~reduces the inference time by up to 29.7x in privacy-preserving model evaluation while keeping the model safe. We share the code at: \url{https://github.com/LiangqinRen/PrivDNN}.

\begin{acks}

Liangqin Ren, Zeyan Liu, Fengjun Li, and Bo Luo were supported in part by NSF IIS-2014552, DGE-1565570, DGE-1922649, and the Ripple University Blockchain Research Initiative. Kaitai Liang was supported in part by the European Union’s Horizon Europe Research and Innovation Programme under Grant No. 101073920 (TENSOR), No. 101070052 (TANGO), and No. 101070627 (REWIRE). The authors would like to thank the anonymous reviewers and the shepherd for their valuable comments and suggestions.

\end{acks}
\bibliographystyle{ACM-Reference-Format}
\bibliography{reference}

\appendix

\section{Glossary}

The following table summarizes the notations used in the paper. 

\begin{table}[ht]
    \centering
    \caption{Notations used in the paper.}\label{tab:notations}
    \begin{tabular}{m{0.11\linewidth} | m{0.80\linewidth}}
    \hline
    Symbol & Notation \\
    \hline
    $A_{o}$ & The classification accuracy of the original deep learning model.  First defined in Section \ref{subsec:privdnnoverview}. \\\hline
    $A_{s}$ & The classification accuracy for authorized clients, who can get decrypted FHE neuron outputs (demonstrated in Figure \ref{fig:separate_remove} (b)). First defined in Section \ref{subsec:objectives}.\\\hline
    $A_{r}$ & The classification accuracy for unauthorized clients who cannot obtain decrypted FHE neuron outputs (Figure \ref{fig:separate_remove} (c)). First defined in Section \ref{subsec:objectives}. \\ \hline
    $A_{rec}$ & The classification accuracy achieved by the curious clients in a model recovery attack. First defined in Section \ref{subsec:modelprivacy}.  \\ \hline
    $A_{t}$ & The classification accuracy for models trained from scratch with 1000 samples. First defined in Section \ref{subsec:modelprivacy}.  \\
    \hline
    $s$ & The quality of the core set selection. First defined in Section \ref{subsec:objectives}, Equation \ref{eqa:total_point}.  \\\hline
    $T$ & Model execution time for a batch of input samples.  \\ 
    \hline
    $n_{i,1}$ & The number of core neurons (with encrypted weights) in the $i$th convolution layer. Defined in Section \ref{subsec:coreselection}. \\ \hline
    $n_{i,2}$ & The number of non-core neurons (with plaintext weights) in the $i$th convolution layer. Defined in Section \ref{subsec:coreselection}. \\
    \hline
    $N_e$ & The total number of core neurons (with encrypted weights). First defined in Section \ref{subsec:privdnnoverview}.\\\hline 
    $N$ & The total number of convolution neurons in the DNN. First defined in Section \ref{subsec:privdnnoverview}.\\\hline 
    $ER$ & The DNN encryption ratio, i.e., the proportion of core neurons out of all convolution neurons in the DNN. First defined in Section \ref{subsec:computationexp}. \\\hline 
    \end{tabular}
\end{table}

\section{Sample Recovery Attack}

We adopt two mechanisms, split learning \cite{erdougan2022unsplit} and autoencoder \cite{autoencoder}, to recover the raw pixels of the input images from the output of the intermediate layers of the network. We evaluate an extreme case where all the outputs from layer 2 are known to the attacker, i.e., the curious model owner. Note that such outputs are obfuscated (multiplied by a random value $\epsilon$). To fully mimic the server's capacity that she has the full training samples, we assume the server will attempt to rescale the obfuscated intermediate results, based on her knowledge of her own training samples, to break the client's obfuscation.

\begin{figure}[ht]
    \centerline{\includegraphics[width=1.0\columnwidth]{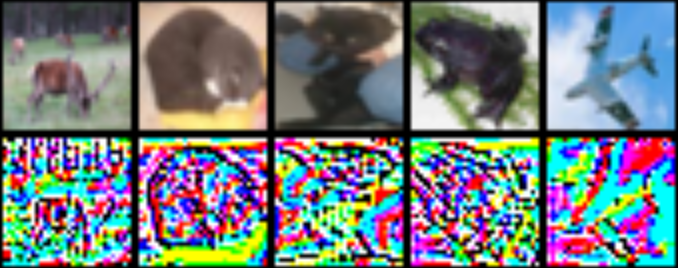}}
    \vspace{-3mm}
    \caption{The model owner's sample recovery attack with split learning \cite{erdougan2022unsplit} when selecting 100\% neurons. }
    \label{fig:data-recovery-unsplit100}\vspace{-2mm}
\end{figure}

In Figure~\ref{fig:data-recovery-unsplit100}, we show the original input images (first row) and the recovery attack results of split learning (second row). As shown, the malicious model owner could not recover any meaningful image. 

\begin{figure}[ht]
    \centerline{\includegraphics[width=1.0\columnwidth]{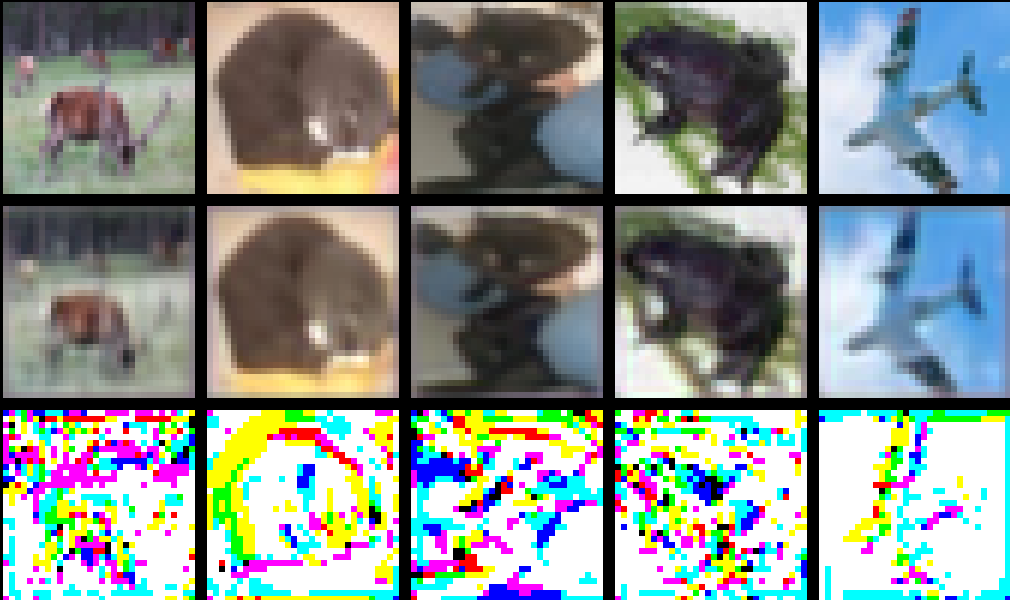}}
    \vspace{-3mm}
    \caption{The model owner's sample recovery attack with an autoencoder \cite{autoencoder} when selecting 100\% neurons.}
    \label{fig:data-recovery-autoencoder100}\vspace{-2mm}
\end{figure}

In Figure~\ref{fig:data-recovery-autoencoder100}, we show the original input images (first row), images recovered by the autoencoder (second row) from unobfuscated intermediate results, and the recovery attack results of the autoencoder (third row) from obfuscated intermediate results.
Even with a well-trained autoencoder, the server cannot recover any meaningful image due to the protection of obfuscation. We also like to note that, in real-world attacks, the model owner is unlikely to have the entire layer 2 encrypted, hence, her capability is even weaker than the example attacks in this section. 

\section{Datasets and Model Structures}\label{apdx:datasets}

\setlength{\textfloatsep}{20.0pt plus 2.0pt minus 4.0pt}
\begin{table}[t]
  \centering
  \caption{Datasets and models used in the experiments.}
  \label{tab:datasets-full}\vspace{-3mm}
  \setlength\tabcolsep{1.9pt}
  \begin{tabular}{c c c c c c}
    \hline
     & M & E & G & C & T \\
    \hline
    Complexity  & Low & Medium & Medium & High & High \\
    Image size  & 28$\times$28 & 28$\times$28 & 32$\times$32 & 32$\times$32 & 64$\times$64 \\
    Categories  & 10 & 26 & 43 & 10 & 200\\
    Train size & 60000 & 124800 & 39209 & 50000 & 100000\\
    Test size & 5000 & 10400 & 6315 & 5000 & 10000\\
    DNN Model & LeNet-5 & LeNet-5 & AlexNet & VGG16 & ResNet18 \\
    \hline
    Optimizer   & Adam & Adam & SGD & SGD & SGD \\
    Learning rate  & 1e-3 & 3e-3 & 5e-2 & 5e-2 & 5e-2 \\
    Scheduler & \multicolumn{5}{c}{CosineAnnealingLR} \\
    Accuracy($A_o$) & 99.36\% & 93.08\% & 93.51\% & 90.22\% & 72.00\%\\
    \hline
  \end{tabular}
  \begin{minipage}{8.25cm}
  \vspace{0.1cm}
  \small  {\textbf{1.} M: MNIST \cite{MNIST_dataset}, E: EMNIST \cite{EMNIST}, G: GTSRB \cite{GTSRB}, C: CIFAR10 \cite{CIFAR10_dataset}, T: Tiny-ImageNet \cite{tinyimagenet} \textbf{2.} We adopt top-5 accuracy for Tiny-ImageNet.}
  \end{minipage}\vspace{-0mm}
\end{table}

\begin{table}[t]
  \centering
  \caption{The structure of the LeNet-5 classifier for MNIST.}
  \label{tab:mnist_model}\vspace{-3mm}
  \begin{tabular}{c    c    c    c    c    c}
    \hline
    Layer & Type & Output & Kernel & Padding & Activation                \\
    \hline
    1 & Conv & 6 & 5$\times$5 & 0 & - \\
    2 & AvePool & 6 & 2$\times$2 & - & Square \\
    3 & Conv & 16 & 5$\times$5 & 0 & - \\
    4 & AvePool & 16 & 2$\times$2 & - & Square \\
    5 & FC & 120 & - & - & ReLU/Square \\
    6 & FC & 84 & - & - & ReLU/Square \\
    7 & Softmax & 10 & - & - & - \\
    \hline 
  \end{tabular}
\end{table}

\begin{table}[t]
  \centering
  \caption{The structure of the LeNet-5 classifier for EMNIST.}
  \label{tab:emnist_model}\vspace{-3mm}
  \begin{tabular}{c    c    c    c    c    c}
    \hline
    Layer & Type & Output & Kernel & Padding & Activation                \\
    \hline
    1 & Conv & 10 & 5$\times$5 & 0 & - \\
    2 & AvePool & 10 & 2$\times$2 & - & Square \\
    3 & Conv & 20 & 5$\times$5 & 0 & - \\
    4 & AvePool & 20 & 2$\times$2 & - & Square \\
    5 & FC & 120 & - & - & ReLU \\
    6 & FC & 84 & - & - & ReLU \\
    7 & Softmax & 27 & - & - & - \\
    \hline 
  \end{tabular}
\end{table}

We adopt five popular benchmarking datasets for machine learning applications, as summarized in Table \ref{tab:datasets-full}. They are briefly introduced as follows:

\noindent $\bullet$ \textbf{MNIST}. The MNIST dataset \cite{MNIST_dataset} contains grayscale images of handwritten digits from 0 to 9. It is a classic dataset for handwritten character recognition. MNIST was derived from a larger dataset, NIST Special DB 19, which also contains uppercase and lowercase letters. We adopt the classic LeNet-5 classifier \cite{LeNet-5} for MNIST. 
  
\noindent  $\bullet$  \textbf{EMNIST}. The Extended MNIST (EMNIST) \cite{EMNIST} is a NIST variant for challenging classification tasks while sharing the same image structure and parameters as MNIST. We use the EMNIST Letters dataset, which contains 26 balanced classes of English letters. To get better accuracy, we modify LeNet-5 by increasing the first layer channels from 6 to 10 and the second layer channels from 16 to 20.

\noindent  $\bullet$  \textbf{GTSRB}. The German Traffic Sign Recognition Benchmark \cite{GTSRB} is an RGB dataset with 43 different types of traffic signs. We resize all the images to $32{\times}32$ pixels in size. We adopt the AlexNet model \cite{AlexNet} and follow the solution in \cite{AlexNet_Kernel} to reduce the kernel size to 5 and 3 for the first two layers, respectively, to fit GTSRB.

\noindent  $\bullet$  \textbf{CIFAR10}. The Canadian Institute For Advanced Research 10 (CIFAR10) dataset \cite{CIFAR10_dataset} is an RGB image dataset with 10 classes, such as automobile, bird, cat, etc. We adopt a more complex model, VGG16 \cite{VGG11}, for CIFAR10.

{\color{black}\noindent  $\bullet$  \textbf{Tiny-ImageNet}. Tiny-ImageNet \cite{tinyimagenet} is a modified subset of the original ImageNet \cite{deng2009imagenet}. It contains 200 different classes of $64{\times}64$ colored images. We adopt ResNet18 \cite{he2016deep} to classify it.}

Next, we present the detailed architectures of the deep learning models used in the experiments: the two LeNet-5 networks for MNIST and EMNIST, respectively, the AlexNet network for GTSRB, the VGG16 network for CIFAR10, and the ResNet18 for Tiny-ImageNet.

\begin{table}[t]
  \centering
  \caption{The structure of the modified AlexNet for the GTSRB dataset.}
  \label{tab:gtsrb_model}\vspace{-3mm}
  \begin{tabular}{c    c    c    c    c    c}
    \hline
    Layer & Type & Output & Kernel & Padding & Activation                \\
    \hline
    1 & Conv & 96 & 5$\times$5 & 1 & - \\
    2 & AvePool & 96 & 2$\times$2 & - & Square \\
    3 & Conv & 256 & 3$\times$3 & 1 & - \\
    4 & BN & 256 & - & - & - \\
    5 & MaxPool & 256 & 2$\times$2 & - & ReLU \\
    6 & Conv & 384 & 3$\times$3 & 1 & ReLU \\
    7 & Conv & 384 & 3$\times$3 & 1 & ReLU \\
    8 & Conv & 256 & 3$\times$3 & 1 & - \\
    9 & MaxPool & 256 & 2$\times$2 & - & ReLU \\
    10 & FC & 512 & - & - & - \\
    11 & Dropout & 512 & - & - & ReLU \\
    12 & FC & 128 & - & - & - \\
    13 & Dropout & 128 & - & - & ReLU \\
    14 & Softmax & 27 & - & - & - \\
    \hline 
  \end{tabular}
\end{table}

\begin{table}[t]
  \centering
  \caption{The structure of the VGG16 model used for CIFAR10}
  \label{tab:cifar10_model}\vspace{-3mm}
  \begin{tabular}{c    c    c    c    c    c}
    \hline
    Layer & Type & Output & Kernel & Padding & Activation \\
    \hline
    1 & Conv & 64 & 3$\times$3 & 1 & Square \\
    2 & Conv & 64 & 3$\times$3 & 1 & - \\
    3 & BN & 64 & - & - & - \\
    4 & MaxPool & 64 & 2$\times$2 & - & ReLU \\
    5 & Conv & 128 & 3$\times$3 & 1 & - \\
    6 & BN & 128 & - & - & ReLU \\
    7 & Conv & 128 & 3$\times$3 & 1 & - \\
    8 & BN & 128 & - & - & - \\
    9 & MaxPool & 128 & 2$\times$2 & - & ReLU \\
    10 & Conv & 256 & 3$\times$3 & 1 & - \\
    11 & BN & 256 & - & - & ReLU \\
    12 & Conv & 256 & 3$\times$3 & 1 & - \\
    13 & BN & 256 & - & - & ReLU \\
    14 & Conv & 256 & 3$\times$3 & 1 & - \\
    15 & BN & 256 & - & - & - \\
    16 & MaxPool & 256 & 2$\times$2 & - & ReLU \\
    17 & Conv & 512 & 3$\times$3 & 1 & - \\
    18 & BN & 512 & - & - & ReLU \\
    19 & Conv & 512 & 3$\times$3 & 1 & - \\
    20 & BN & 512 & - & - & ReLU \\
    21 & Conv & 512 & 3$\times$3 & 1 & - \\
    22 & BN & 512 & - & - & - \\
    23 & MaxPool & 512 & 2$\times$2 & - & ReLU \\
    24 & Conv & 512 & 3$\times$3 & 1 & - \\
    25 & BN & 512 & - & - & ReLU \\
    26 & Conv & 512 & 3$\times$3 & 1 & - \\
    27 & BN & 512 & - & - & ReLU \\
    28 & Conv & 512 & 3$\times$3 & 1 & - \\
    29 & BN & 512 & - & - & - \\
    30 & MaxPool & 512 & 2$\times$2 & - & ReLU \\
    31 & FC & 512 & - & - & - \\
    32 & Dropout & 512 & - & - & ReLU \\
    33 & FC & 512 & - & - & - \\
    34 & Dropout & 512 & - & - & ReLU \\
    35 & Softmax & 10 & - & - & - \\
    \hline 
  \end{tabular}
\end{table}

\begin{table}[t]
  \centering
  \caption{The structure of the ResNet18 model used for Tiny-ImageNet}
  \label{tab:tinyimagenet_model}\vspace{-3mm}
  \begin{tabular}{c    c    c    c    c    c}
    \hline
    Layer & Type & Output & Kernel & Padding & Activation \\
    \hline
    1 & Conv & 64 & 3$\times$3 & 1 & - \\
    2 & BN & 64 & - & - & Square \\
    3 & Conv & 64 & 3$\times$3 & 1 & - \\
    4 & BN & 64 & - & - & ReLU \\
    5 & Conv & 64 & 3$\times$3 & 1 & - \\
    6 & BN & 64 & - & - & ReLU \\
    7 & Shortcut & 64 & - & - & - \\
    8 & Conv & 128 & 3$\times$3 & 1 & - \\
    9 & BN & 128 & - & - & ReLU \\
    10 & Conv & 128 & 3$\times$3 & 1 & - \\
    11 & BN & 128 & - & - &  \\
    12 & Shortcut & 128 & - & - & - \\
    13 & Conv & 128 & 3$\times$3 & 1 & - \\
    14 & BN & 128 & - & - & ReLU \\
    15 & Conv & 128 & 3$\times$3 & 1 & - \\
    16 & BN & 128 & - & - &  \\
    17 & Conv & 256 & 3$\times$3 & 1 & - \\
    18 & BN & 256 & - & - & ReLU \\
    19 & Conv & 256 & 3$\times$3 & 1 & - \\
    20 & BN & 256 & - & - &  \\
    21 & Shortcut & 256 & - & - & - \\
    22 & Conv & 256 & 3$\times$3 & 1 & - \\
    23 & BN & 256 & - & - & ReLU \\
    24 & Conv & 256 & 3$\times$3 & 1 & - \\
    25 & BN & 256 & - & - &  \\
    26 & Conv & 512 & 3$\times$3 & 1 & - \\
    27 & BN & 512 & - & - & ReLU \\
    28 & Conv & 512 & 3$\times$3 & 1 & - \\
    29 & BN & 512 & - & - &  \\
    30 & Shortcut & 512 & - & - & - \\
    31 & Conv & 512 & 3$\times$3 & 1 & - \\
    32 & BN & 512 & - & - & ReLU \\
    33 & Conv & 512 & 3$\times$3 & 1 & - \\
    34 & BN & 512 & - & - &  \\
    35 & Softmax & 200 & - & - & - \\
    \hline 
  \end{tabular}
\end{table}

\end{document}